# Antenna Array Beamforming Based on a Hybrid Quantum Optimization Framework

Shuai Zeng

Chongqing University of Posts and Telecommunications, Chongqing 400065, China

(e-mail: zengshuai@cqupt.edu.cn)

**Abstract**

This paper proposes a hybrid quantum optimization framework for large-scale antenna-array beamforming with jointly optimized discrete phases and continuous amplitudes. The method combines quantum-inspired search with classical gradient refinement to handle mixed discrete-continuous variables efficiently. For phase optimization, a Gray-code and odd-combination encoding scheme is introduced to improve robustness and avoid the complexity explosion of higher-order Ising models. For amplitude optimization, a geometric spin-combination encoding and a two-stage strategy are developed, using quantum-inspired optimization for coarse search and gradient optimization for fine refinement. To enhance solution diversity and quality, a rainbow quantum-inspired algorithm integrates multiple optimizers for parallel exploration, followed by hierarchical-clustering-based candidate refinement. In addition, a double outer-product method and an augmented version are proposed to construct the coupling matrix and bias vector efficiently, improving numerical precision and implementation efficiency. Under the scoring rules of the 7th National Quantum Computing Hackathon, simulations on a 32-element antenna array show that the proposed method achieves a score of 461.58 under constraints on near-main-lobe sidelobes, wide-angle sidelobes, beamwidth, and optimization time, nearly doubling the baseline score. The proposed framework provides an effective reference for beamforming optimization in future wireless communication systems.

# 1 Introduction

Since its concept was proposed, multiple-input multiple-output (MIMO) antenna-array technology has rapidly developed into a cornerstone and standard configuration of modern wireless communication systems [1]. By deploying multiple antennas at both the transmitter and receiver and fully exploiting the spatial degrees of freedom of wireless channels, MIMO antenna arrays overcome the bottlenecks of scarce spectrum resources and harsh transmission environments faced by traditional single-antenna systems, thereby delivering the two major gains of spatial multiplexing and spatial diversity. With the global large-scale deployment of 5G networks and the deepening progress of 6G research, the demand for data traffic and system performance continues to rise year by year. Forecast data from the International Telecommunication Union (ITU) indicate that global mobile data traffic will reach one zettabyte per month by 2030, nearly 30 times the 2020 level [2]. This exponential growth poses further challenges to wireless communication systems: achieving higher transmission rates, broader coverage, and more reliable connectivity under limited spectrum resources. In this context, beamforming technology has attracted extensive attention and research. By precisely controlling the radiation pattern of an antenna array and focusing electromagnetic energy toward target users, beamforming is expected to become a core solution for breaking the capacity bottleneck of conventional communication systems. This directional transmission mechanism not only greatly improves signal reception quality, but also reshapes cellular-network architecture through spatial isolation, allowing adjacent cells to adopt full frequency reuse and thereby multiplying conventional spectrum-reuse efficiency. In other words, without increasing spectrum resources, beamforming can exploit the interference and superposition of electromagnetic waves to realize directional energy transmission in three-dimensional space and substantially enhance the communication link between base stations and terminals. In the evolution of mobile communications across generations, beamforming has gradually become one of the most important enabling technologies and research directions for future mobile communication systems.

At the engineering implementation level, however, beamforming still faces many challenges. First, the beamforming problem of antenna arrays is essentially a high-dimensional nonconvex optimization problem whose complexity grows exponentially with the number of antennas. As system scale expands, the number of parameters requiring joint optimization increases dramatically. For example, current massive MIMO systems may contain dozens or even hundreds of independent antenna elements, and each element has multiple tunable parameters such as phase and amplitude. As a result, MIMO beamforming is effectively transformed into a search task in a high-dimensional combinatorial space whose solution space grows exponentially with the number of antennas, making it a typical NP-hard problem [3]. With larger arrays, conventional optimization methods increasingly fail to satisfy practical engineering requirements. Second, the real-time requirement of beamforming constitutes a second severe challenge. The dynamic nature of modern mobile-communication scenarios requires systems to respond quickly to channel changes: in high-mobility environments, base stations must redirect beams within a very short time to maintain stable connections; in dense-user scenarios, systems must adjust multi-beam coordination in real time to avoid interference. Under such stringent real-time requirements, traditional iterative optimization algorithms face fundamental limitations due to insufficient convergence speed. In particular, when the optimization time exceeds the channel coherence time, the optimization result becomes completely invalid, and beamforming loses its meaning. In addition, beamforming involves inherent conflicts among multiple optimization objectives. It usually needs to satisfy several mutually restrictive metrics at the same time, such as minimizing the main-beam width, maximizing sidelobe suppression, and strictly controlling pointing accuracy. These objectives are intrinsically contradictory: compression of the main beam is usually accompanied by elevated sidelobes, while strict control of pointing deviation restricts the feasible solution space. More importantly, these performance indicators are nonlinearly coupled. In millimeter-wave systems, a slight pointing deviation may cause a significant capacity loss, while insufficient sidelobe suppression may lead to inter-cell interference. Such multi-objective optimization gives rise to a complex Pareto frontier whose solution is far more difficult than that of single-objective optimization.

Quantum annealing has recently shown promise in solving electromagnetic problems because it can obtain high-quality solutions within a relatively short optimization time [4-6]. This idea has been further applied to beamforming optimization for reconfigurable intelligent surfaces. Reference [7] introduced and analyzed an Ising-model-based design method for reconfigurable intelligent surfaces, in which the beamforming Hamiltonian characterizes the total scattered power in a given direction. Case studies on 1-bit and 2-bit pure-phase beamforming for small-scale reconfigurable intelligent surfaces confirmed the effectiveness of quantum annealing [8]. However, although quantum annealing is promising for combinatorial optimization, its platforms face inherent scalability bottlenecks when handling fully connected problems such as MIMO beamforming. Quantum-annealing hardware is limited by qubit connectivity and often requires multiple physical qubits to represent a single fully connected spin variable [8]. For example, although the D-Wave platform integrates more than several thousand qubits, the actual number of fully connected variables it can effectively process is still limited to only a few hundred [9]. Therefore, current hardware is still unable to directly handle large-scale MIMO beamforming optimization. Inspired by the basic principles of quantum annealing, recent studies such as [10-14] have adopted a quantum-inspired method called simulated bifurcation (SB), together with its variants, to solve beamforming problems for large-scale antenna arrays and have achieved strong performance. These algorithms simulate the adiabatic evolution of classical nonlinear oscillators on classical computers, map nonlinear quantum models into a classical framework, and solve the corresponding equations of motion, thereby providing a feasible new route for combinatorial optimization problems, including beamforming optimization of large-scale pure-phase-quantized MIMO arrays, that are difficult for traditional classical computing. Experimental results show that this unique combination of quantum principles and classical computing fully exploits quantum-inspired characteristics, significantly improving both solving speed and solution quality [4], effectively addressing the scalability challenge brought by the complex energy landscape of large-scale MIMO systems and thereby ensuring robust and scalable solutions for practical applications.

Building on the existing literature, this study investigates the beamforming problem and develops a new algorithm. To address the difficult joint optimization of discrete phase quantization and continuous amplitude control in large-scale MIMO antenna-array beamforming, an antenna-array beamforming algorithm based on a hybrid optimization framework combining quantum-inspired methods with classical gradients is proposed. The framework deeply integrates the efficient search capability of quantum-inspired algorithms in discrete spaces with the fine adjustment capability of classical gradient algorithms in continuous spaces, thereby achieving coordinated optimization of discretely quantized phases and continuous amplitudes. On the one hand, this hybrid framework overcomes the limitation of purely quantum-inspired methods in handling continuous variables; on the other hand, it alleviates the tendency of traditional gradient optimization to become trapped in local optima of discrete phases, thus providing a useful solution for beamforming under complex constraints.

The main contributions of this paper are as follows:

(1) An antenna-array beamforming algorithm based on a hybrid optimization framework combining quantum-inspired methods and classical gradients is proposed

The algorithm effectively integrates the efficient search capability of quantum-inspired algorithms in the discrete-phase space with the fine adjustment advantage of classical gradient optimization in the continuous-amplitude space, enabling joint optimization of discrete phases and continuous amplitudes and overcoming the limitations of conventional single-paradigm methods in mixed-variable optimization.

(2) A phase-encoding scheme based on Gray codes and odd-combination products is designed

Through Gray-code transformation, odd-subset feature extraction, and linear mapping, discrete phase states are accurately mapped into the spin-variable space, improving encoding robustness and avoiding the exponential complexity growth of higher-order Ising models, thereby providing a universal theoretical foundation for phase optimization at arbitrary bit precision.

(3) An amplitude-encoding scheme based on geometric combinations of spin variables is designed

After phase optimization by the quantum-inspired algorithm, the proposed method further encodes and solves the amplitude by means of an Ising model. The amplitude value is encoded through a geometric progression of the lengths represented by different spin variables. Compared with either no encoding or equal-length allocation among spin variables, this scheme can approximate the optimized amplitude with higher precision without increasing the model order.

(4) A rainbow quantum-inspired algorithm and a candidate-solution refinement scheme based on hierarchical clustering are proposed

This optimization scheme selectively integrates seven optimizers, namely ballistic simulated bifurcation (BSB), discrete simulated bifurcation (DSB), simulated coherent Ising machine (SimCIM), local quantum annealing (LQA), chaotic amplitude control (CAC), chaotic feedback control (CFC), and noisy mean-field annealing (NMFA), to explore the solution space in parallel, and then refines candidate solutions through a hierarchical-clustering-based algorithm. In this way, it both breaks through the search bias and limitations of a single algorithm to improve solution diversity and compresses the candidate-solution set to retain more effective solutions and improve computational efficiency.

(5) A double outer-product method and an augmented double outer-product method are designed to construct the coupling matrix and bias vector of the objective function

For the coupling matrix of the optimization objective in phase encoding, as well as the coupling matrix and bias vector of the optimization objective in amplitude encoding, appropriate tensor operations are used to construct these quantities conveniently. Compared with general element-wise computation, this method simplifies the calculation process, improves computational efficiency, and enhances numerical stability. It is also compatible with phase and amplitude encodings at arbitrary bit widths, improving code encapsulation and program design.

(6) Exploration of higher-order Ising-model solving

Relevant chapters of the paper discuss the difficulty of solving higher-order models after converting the objective function into an Ising model, the possibility and tradeoffs of exchanging more spin-variable bits for lower dimensionality, and indirect dimensionality-reduction ideas and methods implemented through self-developed utility functions based on SymPy and PyQUBO.

The remainder of this paper is organized as follows. Chapter 2 builds and analyzes the system model by first introducing the basic model of large-scale antenna arrays, then describing the specific beamforming requirements of the competition problem, and finally formulating the problem as a mixed-integer nonlinear program and analyzing its complexity. Chapter 3 focuses on algorithm design by first presenting the basic procedure of the hybrid framework combining quantum-inspired and classical gradient optimization, and then detailing the innovations and design details of the hybrid framework, phase and amplitude encoding schemes, the rainbow quantum-inspired algorithm, and candidate-solution refinement. Chapter 4 presents numerical experiments and analysis by first introducing the scoring criteria and test environment, then comparing the proposed algorithm with baseline algorithms under multiple metrics, and finally describing the overall technical evolution and online scores. Chapter 5 concludes the paper and outlines future work.

## 2 System Model and Analysis

### 2.1 Basic Model

The core challenge of large-scale MIMO antenna-array beamforming lies in achieving precise spatial control of electromagnetic energy through coordinated optimization over a high-dimensional parameter space. Consider a uniform planar array composed of $M \times N$ antenna elements, with each element located at a specific position $(x_m, y_n, 0)$ in a Cartesian coordinate system. Here $x_m = (m-1)d_x$ denotes the coordinate of the element in row $m$ along the $x$ axis, and $d_x$ is the element spacing in the $x$ direction, usually set to half a wavelength $\lambda/2$ to avoid grating lobes; $y_n = (n-1)d_y$ denotes the coordinate of the element in column $n$ along the $y$ axis, and $d_y$ is the element spacing in the $y$ direction. The radiation characteristic of each antenna element is fully described by the complex excitation coefficient $w_{mn} = a_{mn}e^{j\phi_{mn}}$, where the amplitude component $a_{mn} \in R^+$ is a continuously adjustable parameter used to control the radiated power of the element and is usually constrained to the normalized interval $[0,1]$ in engineering practice to guarantee the linear operating range of the power amplifier; the phase component $\phi_{mn}$ controls the interference phase of the electromagnetic wave. Limited by the quantization precision of hardware phase shifters, the phase is usually discretized, and each discrete phase value corresponds to a specific control state of the digital phase shifter. When the electromagnetic wave propagates in the far field, the total field strength at observation point $(\theta, \phi)$ is

accurately described by the array factor:

$$AF(\theta, \phi) = \sum_{m=1}^{M} \sum_{n=1}^{N} a_{mn} \exp(j\phi_{mn}) \exp\left(j \frac{2\pi}{\lambda}(x_m \sin\theta \cos\phi + y_n \sin\theta \sin\phi)\right)$$

In the above expression, $\lambda$ is the operating wavelength, whose physical meaning determines the spatial scale of the electromagnetic wave and directly affects the spatial sampling characteristics of the array; the spatial phase-delay term $\exp\left(j\frac{2\pi}{\lambda}(\cdot)\right)$ physically represents the phase lag caused by differences in propagation path length of the wavefront and is determined by the intrinsic geometric structure of the array. The essence of beamforming is to optimize the excitation parameters of $K = M \times N$ elements under strict engineering constraints so that radiated energy is maximally focused in the target direction (the main-beam direction) $(\theta_0, \phi_0)$, as reflected by the peak of directional gain $G(\theta_0, \phi_0) = |AF(\theta_0, \phi_0)|^2$, while parasitic radiation in non-target regions (sidelobe directions) is suppressed.

**2.2 Competition Requirements**

The quantum-inspired algorithm track of the 7th National Quantum Computing Hackathon appropriately simplified the above model and partially quantified it numerically, and then specified concrete beamforming requirements.

Consider a uniform linear array composed of $N$ antenna subarrays. All subarrays are simplified as being equally spaced only along the $z$ axis, with spacing equal to half a wavelength $\lambda/2$ (as shown in Fig. 1). For a linear array, the radiation field is symmetric with respect to the $\phi$ axis. Therefore, the competition problem fixes $\phi = 90°$, and $AF(\theta, \phi)$ is simplified as a function of $\theta$. Thus, in the far field, the radiation intensity in an arbitrary spatial direction $\theta$ is simplified as the squared magnitude of the array factor $F(\theta)$:

$$G(\theta, \phi) = |AF(\theta, \phi)|^2 = |F(\theta)|^2 = E(\theta) \left[\sum_{n=1}^{N} A_n(\theta)\right] \left[\sum_{n=1}^{N} A_n^*(\theta)\right]$$

The key components are defined as follows:

Array factor $A_n(\theta)$: describes the radiation contribution of the nth subarray and is expressed as

$$A_n(\theta) = I_n \exp\{i\pi n \cos\theta\}$$

where $I_n = \beta_n e^{i\alpha_n}$ is the complex excitation coefficient, $\beta_n \in [0,1]$ denotes the amplitude,

and $\alpha_n$ denotes the phase angle.

Element factor $E(\theta)$: characterizes the directional field pattern of a single antenna subarray, and the competition problem fixes it as

$$E(\theta) = 10^{E_{dB}(\theta)/10}, \quad E_{dB}(\theta) = -min\left\{12\left(\frac{\theta - 90°}{90°}\right)^2, 30\right\}$$

This function reaches its maximum at $\theta = 90°$ with value $0dB$, and decays at $\theta = 0°$ and $180°$ to $-30dB$.

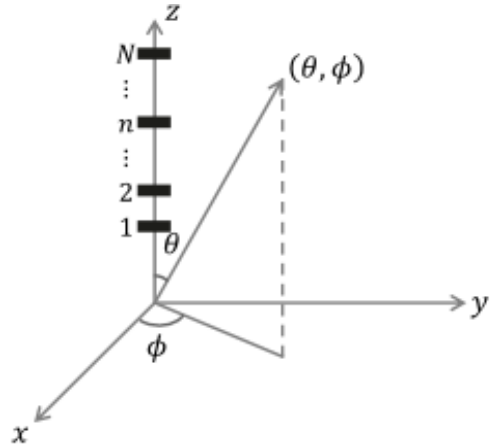

Fig. 1 Schematic of the Antenna Array

Given the main-beam pointing angle $\theta_0 \in [45°, 135°]$, the beamforming algorithm is expected to satisfy four design requirements as well as possible:

(1) Deep suppression in the near-main-lobe region

This requirement arises from interference-management needs in dense-user scenarios. At the target elevation angle $\theta_0$, within the adjacent region $\theta \in [\theta_0 - 30°, \theta_0 + 30°]$, the attenuation of any sidelobe peak power relative to the main-beam gain $|F(\theta_0)|^2$ must be no less than $30dB$. Physically, this constraint is intended to eliminate co-channel interference to nearby users caused by energy leakage at the edge of the main beam. Its mathematical form can be written as:

$$\max_{\theta \in [\theta_0 - 30°, \theta_0 + 30°]} \left(10\lg\frac{|F(\theta)|^2}{|F(\theta_0)|^2}\right) \leq -30$$

The max operation indicates that the worst sidelobe point within this region must be suppressed, which significantly increases the strictness of the constraint. From an electromagnetic perspective, this requirement calls for a highly uniform excitation distribution of the array, because any abnormal fluctuation in element excitation may lead to local sidelobe rise.

(2) Wide-angle sidelobe interference suppression

This requirement serves the high-frequency-reuse architecture of cellular networks. Over the wide angular range $(\theta \in [0°, \theta_0 - 30°) \cup (\theta_0 + 30°, 180°])$, the attenuation of sidelobe power relative to the main-beam peak must be at least $15dB$. The purpose of this constraint is to block long-range interference from the base station to users outside the service area, especially users in neighboring cells. Its mathematical form can be written as:

$$\max_{\theta \in [0°, \theta_0 - 30°) \cup (\theta_0 + 30°, 180°]} \left(10 \lg \frac{|F(\theta)|^2}{|F(\theta_0)|^2}\right) \leq -15$$

Its engineering significance lies in reducing inter-cell interference and improving the signal-to-noise ratio of edge users. Satisfying this constraint usually requires introducing a specific amplitude-tapering strategy into excitation-distribution design, although part of the aperture efficiency is often sacrificed.

(3) Main-beam width compression

This requirement is typically motivated by the need for high-precision spatial resolution in millimeter-wave communications. The power beamwidth of the main beam $3dB$ must be controlled within $8°$, that is, the main-beam width $W$ must satisfy:

$$W = \theta_2' - \theta_1' \leq 8°, \quad |F(\theta_1')|^2 = |F(\theta_2')|^2 = 10^{-3} \cdot max|F(\theta)|^2$$

The physical basis of this metric is the Rayleigh resolution limit in antenna theory. The main-beam width directly determines the spatial multiplexing capability of the system. For example, in the millimeter-wave band, narrow beams can significantly improve user isolation and spectrum efficiency.

(4) Main-beam pointing deviation control

This requirement is a key performance indicator of beamforming technology and is used to control the exact pointing of the main beam. It requires the direction of maximum beam gain to remain strictly aligned with the predetermined target direction $\theta_0$, thereby avoiding unintended interference radiation. Its mathematical expression can be written as:

$$\left|\arg \max_{\theta} |F(\theta)|^2 - \theta_0\right| \leq 1°$$

## 2.3 Problem Formulation and Analysis

As can be seen from the competition description, there are conflicts between the design requirements and the objective of maximizing main-beam gain. Compressing the main-beam width usually requires increasing the effective aperture of the array or adjusting the excitation distribution, but this often raises the sidelobe level. Conversely, strict sidelobe-suppression requirements, such as near-region $30dB$ suppression, tend to force energy to concentrate into the main beam and may sacrifice the beamwidth metric. A more fundamental challenge comes from the mixed nature of the optimization variables: the amplitude $\beta_n$ must be continuously adjusted in the analog RF chain to control the distribution of radiated energy, whereas the phase $\alpha_n$ is usually limited by digital phase-shifter hardware and can only take discrete states while controlling wavefront interference.

Therefore, the beamforming problem in the competition can be formulated as a mixed-integer nonlinear program, whose mathematical expression is

$$\max_{\alpha_n,\beta_n} |F(\theta_0)|^2 = \max_{\alpha_n,\beta_n} \left| E(\theta_0) \sum_{n=1}^{N} \beta_n e^{i\alpha_n} e^{i\pi n \cos\theta_0} \right|^2$$

$$\text{s.t. } C_1: \max_{\theta \in [\theta_0-30°, \theta_0+30°]} \left( 10\lg \frac{|F(\theta)|^2}{|F(\theta_0)|^2} \right) \leq -30$$

$$C_2: \max_{\theta \in [0°, \theta_0-30°) \cup (\theta_0+30°, 180°]} \left( 10\lg \frac{|F(\theta)|^2}{|F(\theta_0)|^2} \right) \leq -15$$

$$C_3: W = \theta_1' - \theta_2' \leq 8°, |F(\theta_1')|^2 = |F(\theta_2')|^2 = 10^{-3} \cdot \max|F(\theta)|^2$$

$$C_4: |\arg\max|F(\theta)|^2 - \theta_0| \leq 1°$$

$$C_5: \alpha_n \in \mathcal{A}, \beta_n \in B \ \forall n \in \{1,2,\dots,N\}$$

This discrete-continuous mixed variable space forms a high-dimensional nonconvex solution space of size $|B|^K \times |\mathcal{A}|^{bK}$, where $B$ and $\mathcal{A}$ denote the value spaces of amplitude $\beta_n$ and phase $\alpha_n$, respectively, and $b$ is the number of discrete phase states. The dimensionality grows exponentially with the number of elements $K$. Traditional optimization methods fail because they cannot handle discrete variables. Common heuristic methods such as genetic algorithms can cope with quantized phases, but when faced with continuous amplitudes they must introduce discretization approximations, which degrade solution quality. Meanwhile, combinatorial explosion makes convergence too slow to satisfy the system time limit. Mathematically, the model exhibits multiple layers of complexity. First, the objective includes the squared magnitude of a complex exponential function, whose Hessian matrix has negative

eigenvalues in the $(\beta_n, \alpha_n)$ space, leading to a multimodal and nonconvex solution landscape. Second, in constraints $C_1$ to $C_4$, the operator $max$ creates gradient jumps at multiple critical points, making purely gradient-based optimization highly prone to convergence failure. In addition, the discrete constraint $C_5$ partitions the solution space into multiple isolated regions. Adjacent regions are connected only through phase jumps, forming a disconnected solution topology.

These characteristics make it difficult for typical solving algorithms to obtain valid feasible solutions within the prescribed time, while the success rate of random search algorithms such as Monte Carlo methods decays exponentially with array scale. This complete optimization model both characterizes the technical requirements of the competition problem and reveals the difficulty and nature of the optimization task, thereby suggesting the solution ideas and framework for the quantum-classical hybrid optimization algorithm proposed in the next chapter.

## 3 Algorithm Design

### 3.1 Basic Procedure

The proposed antenna-array beamforming algorithm based on a hybrid optimization framework is built on a collaborative architecture combining quantum-inspired and classical gradient optimization, and its overall execution follows the paradigm of dual-track advancement, coordinated optimization, and comprehensive evaluation.

The algorithm begins with system initialization, where key parameters such as the target pointing angle, the number of phase-quantization bits, and the amplitude-optimization flag are loaded to establish the electromagnetic base model of the antenna array. The algorithm then performs phase encoding, whose details are given later. After the phase encoding is determined, the algorithm proceeds into two branches: the quantum-inspired optimization branch and the classical gradient optimization branch.

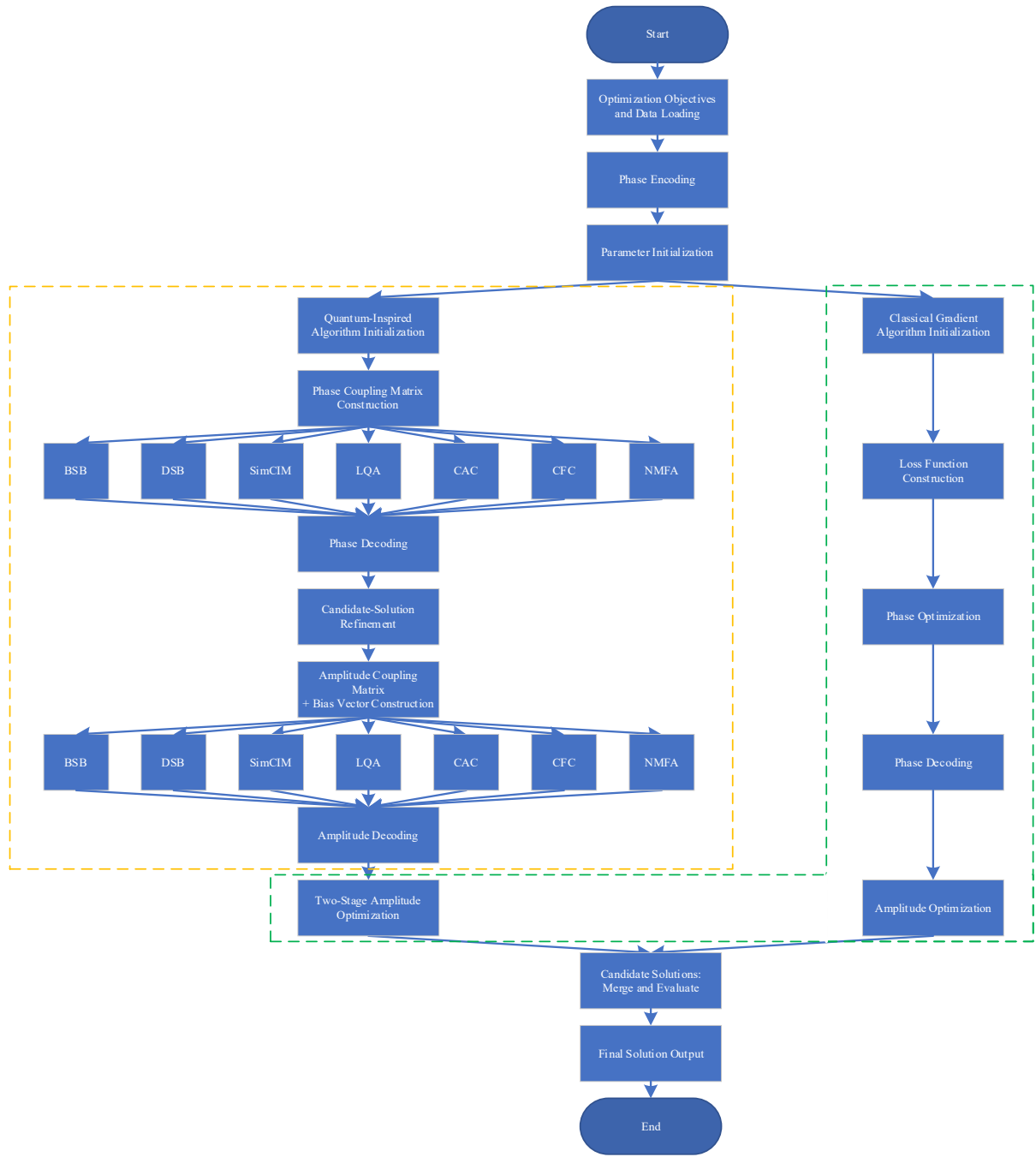

**Fig. 2 Overall Flow of the Algorithm**

In the quantum-inspired optimization branch, the first core task is to solve the discrete phase optimization problem. The algorithm constructs the main-beam gain matrix and the sidelobe-suppression matrix, and then combines them through weighting coefficients to form the overall phase coupling matrix of the optimization objective. The matrix elements quantify the radiation interference effects among antenna elements: the diagonal terms enhance energy focusing in the target direction, while the off-diagonal terms suppress interference at specific angles. The algorithm then invokes seven quantum-inspired solvers in parallel, namely ballistic simulated bifurcation (BSB), discrete simulated bifurcation (DSB), simulated coherent Ising machine (SimCIM), local quantum annealing (LQA), chaotic amplitude control (CAC), chaotic feedback control (CFC), and noisy mean-field annealing (NMFA). After each solver generates its own candidate-solution set, the spin states are inversely mapped into discrete phase values, namely phase decoding. The candidate solutions are then refined through a hierarchical-clustering-based procedure to compress the candidate set. The second important task in the quantum-inspired branch is amplitude optimization. Unlike phase encoding, amplitude encoding is based on geometric combinations of spin variables and requires not only a coupling matrix but also a bias vector. The subsequent optimization is still performed by the quantum-inspired solvers. The output amplitudes are then further optimized through a second-stage gradient-based refinement. Although this second-stage optimization lies within the quantum-inspired path, the method itself is classical gradient optimization. Finally, the selected phase and amplitude values are aggregated into the candidate pool of the quantum-inspired branch.

In the classical gradient optimization branch, the algorithm uses a gradient-based method as the computational engine. It first randomly initializes phase seeds and converts the discrete phase values into a continuously differentiable representation. It then constructs a loss function and performs optimization with a classical gradient optimizer. Similar to the quantum-inspired branch, the classical branch also performs phase decoding and amplitude optimization after phase optimization. The results of the classical branch are then fused with the candidate pool from the quantum-inspired branch, and the final solution is selected through evaluation. The complete flow is shown in Fig. 2.

### 3.2 Innovations and Design Details

### 3.2.1 Hybrid Framework of Quantum-Inspired and Classical Gradient Optimization

The core innovation of this study is the construction of a hybrid computational framework in which quantum-inspired optimization and classical gradient optimization are deeply coordinated to achieve efficient global optimization of discrete phases and continuous amplitudes by exploiting the complementary strengths of the two paradigms. Traditional methods are limited by a single optimization paradigm. Quantum-inspired algorithms are good at handling combinatorial problems of discrete phases but struggle with continuous amplitudes, whereas gradient methods are strong in continuous-space search but tend to fall into local optima for discrete phases. To overcome this limitation, the proposed framework adopts a dual-path architecture in which quantum-inspired and classical gradient optimization advance in parallel.

In the quantum-optimization path, the phase is first encoded and the coupling matrix of the objective function is constructed, after which phase adjustment is carried out with a quantum-inspired algorithm. The optimization objective is the weighted difference between sidelobe power and main-beam power:

$$\min_{\alpha_n, \beta_n} \lambda P_{\text{sidelobe}} + (1-\lambda) P_{\text{mainlobe}}$$

Here the amplitude $\beta_n$ is temporarily fixed at $\mathbf{1}_N$, and the expressions of main-beam power and sidelobe power are given as follows, where the meaning of each variable is the same as in Chapter 2:

$$P_{\text{mainlobe}} = \left| E(\theta_0) \sum_{n=1}^{N} \beta_n e^{i\alpha_n} e^{i\pi n \cos\theta_0} \right|^2$$

$$P_{\text{sidelobe}} = \sum_{j} w_j \left| E(\theta_j) \sum_{n=1}^{N} \beta_n e^{i\alpha_n} e^{i\pi n \cos\theta_0} \right|^2$$

Phase candidate solutions are then obtained through multiple quantum-inspired solvers. On this basis, the amplitude freedom $\beta_n$ is unfrozen and amplitude optimization is performed. The amplitude optimization in the quantum path adopts a two-stage structure. The first stage is a coarse amplitude adjustment based on a quantum-inspired algorithm to escape local optima. Specifically, amplitude encoding is introduced and the coupling matrix and bias vector of the objective function are constructed to obtain a preliminarily good amplitude value on a global scale. The second stage then refines the amplitude with classical gradient optimization to obtain a precise amplitude value.

In the classical optimization path, the processing flow mainly follows the gradient-optimization example algorithm provided for the competition problem, namely constructing a weighted-ratio loss function based on main-beam and sidelobe powers:

$$\min_{\alpha_n, \beta_n} \frac{P_{\text{sidelobe}}}{\lambda P_{\text{mainlobe}}}$$

On this basis, traditional gradient optimization is then used to optimize both phase and amplitude, and the resulting solutions are injected into the candidate pool of the quantum path for coordinated management. The final output is selected through unified evaluation.

In summary, the proposed algorithm achieves coordinated optimization of discrete and continuous variables and thereby realizes complementary quantum-classical advantages: the quantum-inspired algorithm breaks the combinatorial-explosion difficulty of discrete-phase optimization and escapes local extrema, while the classical gradient method refines continuous amplitudes. Ultimately, the algorithm outputs a high-performance beamforming solution while satisfying the required constraints. In the flowchart of Fig. 2, the orange dashed box indicates the quantum-inspired part of the hybrid framework, and the green dashed box indicates the classical gradient-optimization part.

**3.2.2 Phase-Encoding Scheme**

To effectively generate the objective-function coupling matrix required by the quantum-inspired algorithm, it is necessary to devise a spin-variable encoding scheme that can represent phase effectively. This study proposes several designs to accomplish this task.

First, based on the encoding method in the competition reference paper, we propose an arbitrary-bit phase encoding generated from Gray codes and odd-combination products, referred to below as the Gray-odd-combination encoding. The mathematical principle of this encoding is to construct an accurate mapping between phase states and spin variables using as few encoding bits as possible while maintaining high robustness. The encoding procedure follows three progressive mathematical transformations. First, a Gray-code transformation is applied: through the bitwise formula $gray = i \oplus (i \gg 1)$, the integer index $i \in [0, 2^b - 1]$ is converted into an $b$-bit Gray-code sequence. The key property of this sequence is that adjacent states differ by only one bit, that is, the Hamming distance is always 1, which guarantees smooth and efficient state transitions during quantum optimization. Next, odd-order combination features are extracted from the Gray codes in the half-space. For the half-space Gray-code set, the product of the elements in each odd-sized subset of size $1, 3, \ldots, b$ is computed, forming a combination-feature matrix $2^{b-1} \times 2^{b-1}$ denoted by $S$, whose element is $S_{kj} = \prod_{m \in O_j} G_{km}$. Here $O_j$ denotes the index of the $j$th odd spin-variable subset, and $G_{km}$ is the Ising value of the $m$th bit in row $k$ of the Gray-code matrix, taking values $\pm 1$. Finally,

through solving the linear system $Sc = p$, the optimal encoding-coefficient vector $c$ is determined, where:

$$p = \left[e^{i\phi_0}, e^{i\phi_1}, \ldots, e^{i\phi_{2^{(b-1)}-1}}\right]^T$$

Here $\phi_k = 2\pi k/2^b$ is the target phase vector, whose entries are ideally and uniformly distributed phase points. The final encoding is completed by taking the dot product of the coefficient vector and the spin-variable vector.

The construction process of the encoding is as follows:

(a) Construct the half-space Gray-code matrix

$$G \in \{-1,1\}^{2^{b-1} \times b}$$

Its matrix element $G_{km}$ denotes, for the $k$th Gray-code state at bit position $m$, the Ising value $(k = 0,1,\ldots,2^{b-1}-1)$. This matrix contains only the first half of the original $2^b$ Gray-code states, that is, the half-space Gray codes. It is generated by the following bitwise operation:

$$i_k = k \, (0 \leq k \leq 2^{b-1}-1)$$

$$\text{gray}_k = i_k \oplus \lfloor i_k \gg 1 \rfloor$$

$$G_{km} = \begin{cases} 1 & \text{bit}_m(\text{gray}_k) = 0 \\ -1 & \text{bit}_m(\text{gray}_k) = 1 \end{cases}$$

(b) Construct the odd-subset combination matrix

$$S_{kj} = \prod_{m \in O_j} G_{km} \in \mathbb{R}^{2^{b-1} \times 2^{b-1}}$$

Here $O_j \subseteq \{1,2,\ldots,b\}$ denotes the $j$th odd spin-variable subset; $|O_j|$ is odd $(1,3,5,\ldots,\leq b)$; and the subset index $j = 0,1,\ldots,2^{b-1}-1$ is arranged in binary order.

(c) Construct the target phase vector

The elements of the target phase vector $p \in \mathbb{C}^{2^{b-1}}$ are defined as

$$p_k = \exp\left(i \cdot \frac{k\pi}{2^{b-1}}\right) \quad (k = 0,1,\ldots,2^{b-1}-1)$$

which represents phase points uniformly distributed over the range $[0,\pi]$ on the unit circle in the complex plane, with a total of $2^{b-1}$ points.

(d) Compute the encoding-coefficient vector

The encoding-coefficient vector $c \in \mathbb{C}^{2^{b-1}}$ is the core data of the encoding and is constructed by the following formula:

$$c = S^{-1}p$$

During phase encoding, the phase can be recovered simply by multiplying the spin variables by the corresponding encoding coefficients and summing them.

(e) Phase encoding

$$e^{i\alpha_n} = c \cdot x_n = \sum_{i=0}^{2^{b-1}-1} c_i X_{ni}$$

As noted above, phase encoding is obtained simply by multiplying the spin variables by the corresponding encoding coefficients and summing them. In the above expression, $x_n$ denotes the vector composed of the spin-variable values obtained after phase optimization for the $n$th antenna subarray, and $\alpha_n$ is the phase angle of the $n$th antenna subarray. The vector dot product can also be written as the sum of the products between each encoding-coefficient component $c_i$ and each spin variable $X_{ni}$.

The core theoretical advantage of this encoding scheme lies in its mathematical completeness: through a linear combination of odd-combination products, any discrete phase state on the unit circle of the complex plane can be reconstructed exactly, and the mapping process preserves the second-order interaction form of the Ising model. In other words, in the phase energy function $E(s) = -\sum_{i<j} J_{ij} s_i s_j$, the spin variables $s_i$ remain quadratic, avoiding the computational disaster of exponential complexity growth caused by higher-order terms. In particular, when $b = 1$ and $b = 2$, the scheme degenerates into standard orthogonal encoding; when $b = 3$ and $b = 4$, it generates complete 8-dimensional and 16-dimensional encoding spaces. More generally, when $b = k$, it generates an $2^k$-dimensional space, thereby ensuring low-order consistency between numerical optimization and encoding. This encoding framework based on combinatorial mathematics establishes a universal foundation for arbitrary-bit phase optimization and opens a path for applying quantum-inspired algorithms to discrete constrained optimization.

In practical algorithm implementation, our tests show that in 1-bit and 2-bit scenarios, directly using the proposed Gray-odd-combination encoding performs best, and in some 3-bit cases, implementing the method with a 4-bit Gray-odd-combination encoding yields the best performance. In 4-bit cases, we also constructed a solution that combines one-hot encoding with linear encoding and solves the encoding coefficients through a pseudoinverse matrix, which can avoid performance degradation caused by too many spin variables. Experiments show that this encoding can be slightly better than Gray-odd-combination encoding in some scenarios, but for consistency the program uniformly adopts Gray-odd-combination encoding. In addition, to achieve the best algorithm performance, we precomputed the encoding coefficients offline and stored them as fixed numerical values in the program rather than generating them dynamically during each optimization run. Of course, if needed in future real MIMO systems, this can be replaced at any time with an on-demand dynamic generation mechanism.

### 3.2.3 Rainbow Quantum-Inspired Algorithm and Candidate-Solution Refinement Based on Hierarchical Clustering

The proposed algorithm integrates multiple quantum-inspired techniques to jointly handle discrete phase optimization. This fused paradigm is named the rainbow quantum-inspired algorithm. Its central design idea is to overcome the inherent limitations of any single algorithm through multi-principle collaboration. The architecture executes multiple quantum-inspired optimizers simultaneously. These optimizers share a unified Hamiltonian-energy representation but remain independent in state initialization, parameter configuration, and evolution path, thereby forming complementary exploration strategies, much like different colors of light in a spectrum jointly illuminating the entire solution space. The specific design and functions of the algorithm include:

(1) Rainbow quantum-inspired algorithm and heterogeneous candidate solutions

The core architecture of the rainbow quantum-inspired algorithm lies in integrating multiple complementary quantum optimizers to explore the solution space in parallel, thereby constructing a search mechanism based on collaboration among multiple physical principles.

Specifically, in the quantum-optimization branch of the algorithm, seven algorithms are deployed simultaneously: BSB, DSB, SimCIM, LQA, CAC, CFC, and NMFA. Each of them builds its solving mechanism on a distinct quantum-physical principle. BSB simulates the evolution of spin variables in a ballistic inertial system to search for the lowest-energy point.

DSB is a discrete improvement of BSB that removes continuous-time evolution and uses iterative update equations to simulate discrete jump-like dynamics. SimCIM simulates optical interference and performs parallel iterations of photon intensity in coherent-state space to approach the global optimum. LQA divides the global optimization problem into multiple local regions, performs short-cycle quantum annealing within each region, and then merges and coordinates the local results to form a better solution. CAC controls the amplitude of chaotic variables to conduct a coarse-to-fine search in the solution space and avoid the local-optimum trap of traditional annealing. CFC uses chaotic signals as part of feedback regulation to control the search direction or system state so as to achieve stability or optimization objectives. NMFA improves mean-field annealing by introducing a noise perturbation term, thereby enhancing the ability to escape local optima.

These algorithms run independently under the shared target Hamiltonian framework. Each algorithm maintains its own set of state variables and requires no cross-algorithm communication, enabling a fully decoupled parallel-computing process. Based on their respective strengths, different algorithms generate different initial candidate solutions, forming a large and heterogeneous sampling of the solution space that covers different characteristic regions of the high-dimensional discrete space. This design fully exploits the multi-path exploration capability of quantum-inspired methods. During independent evolution, state updates are completed entirely in local memory without synchronization; only after the final iteration are the candidate solutions submitted to the central processor, which minimizes communication overhead while maintaining computational efficiency. The collaboration among the seven algorithms effectively constructs a virtual quantum-computing network that simulates quantum parallelism on classical hardware and provides an efficient solution for large-scale discrete optimization problems.

(2) Hierarchical clustering and candidate-solution refinement

After multiple quantum-inspired algorithms generate candidate-solution sets, the algorithm enters the second-stage optimization process: candidate-solution refinement. This part is further divided into exact deduplication and approximate-solution merging. Exact deduplication is straightforward: if multiple identical spin-variable vectors appear in the candidate set, only one is retained. If the candidate set is still too large after exact deduplication, hierarchical clustering is then used to merge approximate solutions with similar phase relationships so as to improve the overall efficiency of the algorithm.

The specific algorithm is as follows:

(a) Input parameters

Candidate-solution set: $\mathcal{S} = \{s_1, s_2, \ldots, s_M\}$, $s_i \in \{-1,1\}^K$, $K = N \times b$

(where $M$ denotes the number of candidate solutions, $N$ denotes the number of antenna elements, and $b$ denotes the number of phase-quantization bits)

Number of refined candidate solutions: $m$

(b) Algorithm procedure

Phase decoding:
$$\alpha^i = decode(s_i), \quad i \in (0,1,\ldots,M)$$

Initialize the cluster set:
$$\mathcal{C} \leftarrow \mathcal{C}_1, \ldots, \mathcal{C}_M, \mathcal{C}_i = \alpha^i$$

Construct the distance matrix with $2\pi$-periodicity taken into account:
$$\mathbf{D}_{ij} = \frac{1}{N} \sum_{k=1}^{N} \min(|\alpha_k^i - \alpha_k^j|, 2\pi - |\alpha_k^i - \alpha_k^j|)$$

Main loop for cluster merging: while $|\mathcal{C}| > m$

Find the pair of clusters with the smallest average distance:
$$(p^*, q^*) = \arg\min_{p<q} \left[ \frac{1}{|\mathcal{C}_p||\mathcal{C}_q|} \sum_{\alpha^i \in \mathcal{C}_p} \sum_{\alpha^j \in \mathcal{C}_q} \mathbf{D}_{ij} \right]$$

Merge clusters:
$$\mathcal{C}_{new} = \mathcal{C}_{p^*} \cup \mathcal{C}_{q^*}$$

Update the cluster set:
$$\mathcal{C} \leftarrow \mathcal{C} \setminus \mathcal{C}_{p^*}, \mathcal{C}_{q^*} \cup \mathcal{C}_{new}$$

Update the distance matrix:
$$d_{new,r} = \frac{\mathcal{C}_{p^*} \cdot d_{p^*,r} + \mathcal{C}_{q^*} \cdot d_{q^*,r}}{\mathcal{C}_{p^*} + \mathcal{C}_{q^*}}, \quad \forall r \neq new$$

(c) Select the representative solution, namely the point with the minimum sum of distances to all other points in the cluster, and output it, where $\mathcal{R}$ is the refined result:

$$\text{for } \mathcal{C}_k \in \mathcal{C} \text{ do}$$
$$\alpha_k^* = \arg\min_{\alpha \in \mathcal{C}_k} \sum_{\alpha' \in \mathcal{C}_k} \mathbf{D}_{\text{idx}(\alpha),\text{idx}(\alpha')}$$
$$\mathcal{R} \leftarrow \mathcal{R} \cup \alpha_k^*$$

The basic idea of the above algorithm comes from hierarchical clustering in machine learning. Its core idea is to merge similar candidate solutions bottom-up and finally form a representative solution set. Specifically, each candidate solution is initially treated as a cluster, after which the distances between clusters are computed to form a distance matrix. In each iteration, the closest pair of clusters is merged and the distance matrix is updated until the target number of refined candidate solutions is reached.

Several key points should be noted. First, because the optimization target is the phase angle, the distance matrix is constructed using amplitude values with periodicity $2\pi$ as the distance metric. In other words, the distance between $\alpha$ and $2\pi + \alpha$ is zero, so the calculation must be taken modulo $2\pi$. Second, when clusters are merged, a representative solution must be chosen. In this algorithm, the point with the minimum sum of distances to all points in the cluster is taken as the cluster representative, which can be regarded as an approximate cluster center in the cluster sense. In addition, the number of refined candidate solutions $m$ can be adjusted according to the practical situation: when the candidate set is large and the remaining computation time is short, it can be reduced; conversely, when phase optimization is unsatisfactory, it can be increased appropriately.

### 3.2.4 Amplitude Encoding and Two-Stage Quantum-Classical Optimization

After phase optimization is completed and refined candidate solutions are obtained, the algorithm enters the amplitude-optimization stage, which is executed when the amplitude-optimization flag is True. This stage adopts a two-stage collaborative strategy combining quantum-inspired methods and classical gradients: in the first stage, amplitude is discretely and coarsely adjusted through amplitude encoding and a quantum-inspired algorithm; in the second stage, amplitude is continuously refined through a classical gradient algorithm. This two-stage mechanism both exploits the ability of quantum-inspired methods to escape local optima and takes advantage of the precise adjustment capability of classical gradient methods in continuous spaces, while also improving overall optimization speed and computational efficiency.

The main processes and techniques in the two stages are introduced below.

(1) Amplitude encoding and optimization with a quantum-inspired algorithm

To enable optimization by a quantum-inspired algorithm, the amplitude value must first be mapped into a function of spin variables, namely spin-variable encoding. The encoding scheme is based on a linear sum of spin variables arranged in a geometric progression. Through such a geometric combination, the amplitude value can be expressed accurately, and compared with an equal-increment encoding scheme, numerical precision can be significantly improved without increasing the model order.

Consider the normalized amplitude range $[0,1]$. If each spin variable represents an equal interval, i.e., equal-value encoding, then each of the $b$ spin variables represents a length of $1/b$, and the minimum resolution is $1/b$. If instead each spin variable represents a geometrically scaled interval, then the $b$ spin variables represent lengths of $1/2$, $1/4$, ..., and $1/2^b$, respectively. It can be shown that such a geometric progression of spin variables achieves the finest possible resolution for a given number of spin variables. In fact, this encoding is equivalent to using spin variables to perform a binary encoding over the range $[0,1]$.

Let the encoding vector be $x = (x_0, x_1, \ldots, x_{b-1})$, where $x_k \in \{0,1\}, k = (0,1,\ldots,b-1)$, and let the coefficient vector be $c = (c_0, c_1, \ldots, c_{b-1})$, where $c_k = 1/2^{k+1}, k = (0,1,\ldots,b-1)$. Then the amplitude value $\beta$ can be expressed in the following form, namely binary encoding:

$$\beta = c \cdot x$$

Considering that the spin variables in the Ising model take values $s \in \{+1, -1\}$, the above expression can be mapped by $x_k = (1 - s_k)/2$. The amplitude value $\beta$ is then written as:

$$\beta = \sum_{k=0}^{b-1} c_k (1 - s_k)/2$$

Let the spin-variable vector be $s = (s_0, s_1, \ldots, s_{b-1})$, where $s_k \in \{0,1\}$, $k = (0,1,\ldots,b-1)$, and redefine the coefficient vector as $c = (c_0, c_1, \ldots, c_{b-1})$, where $c_k = 1/2^{k+2}$, $k = (0,1,\ldots,b-1)$. The vector form of the amplitude value $\beta$ is then

$$\beta = c \cdot (\mathbf{1}_b - s)$$

At this point, the amplitude ranges from $\beta_{min} = c_{b-2}$ to $\beta_{max} = 1$, with a minimum resolution of $\Delta\beta_{min} = 1/2^b$. For example, when $b$ is set to 4, the finest amplitude-adjustment precision reaches 0.0625, which is four times better than the 0.25 of equal-value encoding. In the final formulation, the $\boldsymbol{c} \cdot \boldsymbol{s}$ part of the amplitude encoding contributes to the coupling matrix of the quantum-inspired algorithm, while the $\boldsymbol{c} \cdot \boldsymbol{1}_b$ part contributes to the bias term. We construct them using what we call the augmented double outer-product method, which is introduced in detail later.

(2) Classical gradient optimization

After the quantum-inspired algorithm adjusts the amplitude, the algorithm further performs second-stage refinement based on classical gradient optimization. Taking the amplitude output of the previous stage as the starting point, it constructs a differentiable loss function and computes gradient information. Specifically, the amplitude is initialized to the discrete value output by the quantum-inspired stage, and then its continuous degree of freedom $\beta_n \in [0,1]$ is activated as the optimization variable to define a subproblem. Through an adaptive learning-rate mechanism, namely the Adam algorithm, the amplitude is fine-tuned in continuous space so as to overcome the precision limitation of quantum-inspired methods in continuous-variable optimization.

This design perfectly combines the global exploration advantage of quantum-inspired methods with the precision-control ability of classical optimization. It both uses the quantum-inspired algorithm to escape local optima and exploits gradient methods for precise adjustment in continuous space, thereby avoiding the performance loss caused by failure of mixed-variable coordination in traditional methods. It can reduce the computational resources consumed by a single optimization while ensuring that solution quality reaches coverage of the Pareto frontier. Finally, the scores of all amplitude candidate solutions are computed, and the most suitable amplitude value is selected for each phase candidate solution to complete the entire amplitude-optimization process.

### 3.2.5 Constructing the Objective-Function Coupling Matrix and Bias Vector by the Double Outer-Product Method and the Augmented Double Outer-Product Method

The use of the double outer-product method and the augmented double outer-product method to construct the coupling matrix and bias vector of the objective function is essentially an algorithmic shortcut derived from mathematical principles. With only two outer-product operations, this method can generate the coupling matrix and bias vector required by the quantum-inspired algorithm. Compared with element-by-element construction, the double outer-product method realizes global construction of matrices and vectors through tensor outer products. Its core innovations are reflected at three levels. First, at the phase- and amplitude-encoding level, the radiation contribution of each antenna element is modeled as the tensor product of a position factor and an encoding vector. Second, at the level of the coupling matrix and bias vector, vector outer products are used to directly construct the quadratic form of the objective function and the coefficients of the bias term, avoiding the inefficient element-wise accumulation in the example code. Third, at the implementation level, properly calling matrix-operation-related functions makes it possible to realize phase and amplitude encodings at arbitrary bit widths, thereby improving code design, computational efficiency, and numerical stability.

The mathematical principles of the double outer-product method and the augmented double outer-product method are as follows:

(1) Constructing the phase coupling matrix by the double outer-product method

The coupling matrix of the phase-optimization objective is the standard quadratic form required by the quantum-inspired algorithm, i.e., the input format of the quantum optimizer, and can be written as follows, where $s$ denotes the spin-variable vector:

$$E(s) = -s^T J s, \quad s \in \{-1,1\}^{Nb}$$

The diagonal elements of the coupling matrix $J$ are

$$J_{ii} = \sum_{k=1}^{b} |c_k|^2 |e^{j\pi m \cos \theta_0}|^2, \quad (m = \lfloor i/b \rfloor)$$

which represent the intrinsic radiation intensities of the elements, while the off-diagonal elements are

$$J_{ij} = \text{Re}\left(\sum_{k=1}^{b} c_k c_k^* e^{j\pi(m-n)\cos\theta_0}\right), \quad (m = \lfloor i/b \rfloor, n = \lfloor j/b \rfloor)$$

which quantify the interference effects between elements.

To construct this coupling matrix, only the following operations are required. First, give the electric-field vector in the target direction $\theta^0$:

$$\mathbf{e}_{\theta_0} = \begin{bmatrix} e^{j\pi \cdot 1 \cdot \cos \theta_0} \\ e^{j\pi \cdot 2 \cdot \cos \theta_0} \\ \vdots \\ e^{j\pi \cdot N \cdot \cos \theta_0} \end{bmatrix}$$

where $N$ denotes the number of antenna elements, and this $N$-dimensional complex vector characterizes the spatial phase distribution of the electromagnetic wave in the array.

Next, give the coefficient vector of the Gray-odd-combination phase encoding:

$$\mathbf{c} = \begin{bmatrix} c_0 \\ c_1 \\ \vdots \\ c_{b-1} \end{bmatrix}$$

where $b$ denotes the number of phase-quantization bits, and $c_k$ denotes the complex encoding coefficients.

Then perform two simple outer-product operations:

First outer product, forming the $N \times b$ matrix

$$\mathbf{v} = \mathbf{e}_{\theta_0} \otimes \mathbf{c}$$

Second outer product, forming the $Nb \times Nb$ matrix

$$J = Re(\mathbf{v} \otimes \mathbf{v}^H)$$

By using the properties of complex quadratic forms and Hermitian matrices, it can be shown mathematically that taking the real part of the result of these two outer-product operations yields a real symmetric matrix, which is exactly the phase coupling matrix. This matrix is equivalent to the one obtained by scalar-by-scalar computation, which requires repeated multiplication, summation, and taking the real part, or by Einstein summation.

(2) Constructing the amplitude coupling matrix and bias vector by the augmented double outer-product method

From the mathematical expression of amplitude encoding introduced earlier, it can be seen that the encoding of the $n$th antenna amplitude value $\beta_n$ is actually composed of two parts, as shown below:

$$\beta_n = \mathbf{c} \cdot (\mathbf{1}_b - \mathbf{s}_n) = -\mathbf{c} \cdot \mathbf{s}_n + \mathbf{c} \cdot \mathbf{1}_b$$

The first part $-\mathbf{c} \cdot \mathbf{s}_n$ can be made consistent with the phase-encoding form simply by adding a minus sign, but the second part $\mathbf{c} \cdot \mathbf{1}_b$ does not exist in phase encoding and therefore the double outer-product method cannot be applied directly to construct the parameters required by the quantum-inspired algorithm. In fact, the numerical value of the second part eventually forms the bias vector of the quantum-inspired algorithm. In this case, we modify the double outer-product method as follows:

First, construct an augmented coefficient vector $c'$ so that

$$c' = \begin{bmatrix} c'_0 \\ c'_1 \\ \vdots \\ c'_{b-1} \\ c'_b \end{bmatrix} = \begin{bmatrix} -c_0 \\ -c_1 \\ \vdots \\ -c_{b-1} \\ \sum_{i=0}^{b-1} c_i \end{bmatrix}$$

Then, similar to the double outer-product method, construct the $J'$ matrix in the following form:

First outer product, forming the $N \times (b+1)$ matrix

$$v' = \mathbf{1}_N \otimes c'$$

Second outer product, forming the $N(b+1) \times N(b+1)$ matrix

$$J' = v' \otimes v'^T$$

Finally, extract from the matrix $J'$ the coupling matrix $J$ and bias vector $h$ required by the quantum-inspired algorithm: take from $J'$ the $Nb \times Nb$ part, namely the first $Nb$ rows and $Nb$ columns, as $J$; delete from $J'$ its last row $N$, and then take the sum of the last $N$ column, that is, sum over rows, as $h$. This corresponds to the following slicing operation:

$$\begin{cases} J = J'[0:Nb, 0:Nb] \\ h = (\sum J'[Nb+1:])[Nb+1:] \end{cases}$$

The universal advantage of the double outer-product method and the augmented double outer-product method is particularly reflected in their compatibility with arbitrary phase and amplitude encoding schemes. Whether for 1-bit binary phase and amplitude or for fine control at 4-bit precision, and even for coupling matrices and bias vectors at arbitrary bit widths, one only needs to adjust the phase-encoding vector for seamless adaptation, without reconstructing the algorithmic logic. In addition, these methods offer several other advantages: matrix construction is completed in one shot by outer-product operations, eliminating the precision loss caused by pointwise accumulation in the example code; modular design becomes easier because the construction logic of the main beam and sidelobes can be separated, improving code readability and maintainability; and the reuse of precomputed electric-field matrices reduces memory-access frequency.

Especially in multi-bit phase-encoding scenarios, the double outer-product method and the augmented double outer-product method are naturally suitable for joint use with the proposed

Gray-odd-combination phase encoding and geometric spin-combination amplitude encoding. By expanding the encoding dimension, they resolve the difficulty of higher-order interactions. Compared with traditional schemes, they simplify programming statements, are easier to encapsulate, improve numerical precision during computation, and reduce computational-resource consumption.

**3.2.6 Dimensionality Reduction of Higher-Order Ising Models**

During the design of the beamforming algorithm in this study, we systematically explored dimensionality-reduction techniques to handle higher-order Ising models arising in 3-bit and 4-bit phase-encoding scenarios. When confronting higher-order interaction problems, the limitations of commonly used toolkits such as PyQUBO and D-Wave libraries become evident: current mainstream toolkits do not directly support the complex-coefficient operations required by quantum optimization. To overcome this limitation, we attempted to develop a modeling workflow based on SymPy symbolic algebra. First, the full complex-coefficient Ising-model expression is constructed using SymPy's symbolic-processing capability; then a dedicated SymPy2PyQUBO conversion module is written to transform SymPy expressions into the PyQUBO format. This converter adopts a three-layer processing mechanism: in the initialization stage, all spin variables are explicitly declared as real-valued to avoid conflicts with complex types; in the core conversion stage, the real and imaginary parts are handled separately, the imaginary part is realified, and polynomial expansion is used to convert the SymPy symbolic model into a data format recognizable by PyQUBO; in the output stage, PyQUBO's built-in to_qubo() method is automatically invoked to perform efficient and accurate dimensionality reduction and convert the higher-order Hamiltonian into the standard quadratic form. This workflow enables complete modeling of complex interference effects such as third-order and higher interactions that could not otherwise be expressed directly. The coupling matrix and bias vector after reduction can then be properly extracted through a custom PyQUBO2QAIA function so as to match the standard input format of the quantum-inspired algorithm.

However, extensive tests show that in the 32-antenna-array scenario required by the competition problem, the automatic dimensionality-reduction mechanism of PyQUBO causes a sharp increase in auxiliary variables, expanding the actual solution space to several times the size of the original problem and introducing substantial numerical errors in model optimization. Experiments indicate that, in 3-bit scenarios, using a 4-bit Gray-odd-combination encoding, and in 4-bit scenarios using an 8-bit Gray-odd-combination encoding or a scheme combining one-hot encoding with linear encoding, both yield better candidate solutions than higher-order Ising-model dimensionality reduction. Therefore, the dimensionality-reduction scheme for higher-order Ising models was not ultimately incorporated into the final competition program. Although it was not formally deployed in this competition problem, the exploration itself remains meaningful. It reveals the dimensionality barrier currently faced by quantum-inspired algorithms in handling higher-order optimization problems, and this insight directly motivated the phase- and amplitude-encoding schemes described earlier, thereby pointing out the technical direction for the competition algorithm.

### 3.2.7 Parameter Tuning

To further improve optimization quality, a large number of parameters and hyperparameters were explored during algorithm design through Bayesian optimization, grid search, manual adjustment, and related methods. The main aspects include:

(1) Batch size and iteration count of the quantum-inspired algorithm;

(2) The specific algorithms included in the rainbow quantum-inspired algorithm and their number;

(3) Candidate-solution refinement ratio;

(4) Number of spin variables used in amplitude encoding;

(5) Number of iterations for phase optimization and amplitude optimization in the gradient-based algorithm;

(6) Number of phase-encoding bits in the gradient-based algorithm;

(7) Learning rate of the gradient-based algorithm;

(8) The xi and dt parameters of the quantum-inspired algorithm;

(9) The coefficient and relation parameters for the main-beam and sidelobe terms in the objective function, including ratio terms based on division and relation terms based on addition.

### 3.2.8 Other Utility Functions and Explorations

In addition to the program code required by the competition problem, a large number of

auxiliary utility functions were also developed in this study, mainly including:

(1) Training-data generation: scripts for generating local training data that satisfy the competition requirements;

(2) Visualization of phase-encoding effects: a visualization program for conveniently observing the graphical effect of phase encoding;

(3) Generator of Gray-odd-combination encoding coefficients: a coefficient generator that helps produce coefficients for phase-encoding cases. This program runs outside the main algorithm to generate phase-encoding coefficients, and the resulting coefficients are directly written into the algorithm as fixed values to save runtime during actual optimization;

(4) Generator of one-hot and linear encoding coefficients: similar to item (3), used as an alternative and comparative scheme for generating phase-encoding coefficients;

(5) Optimization with equal-value amplitude encoding: used to compare with the proposed geometric amplitude-encoding scheme;

(6) Higher-order dimensionality-reduction utility functions: utility functions such as SymPy2PyQUBO and PyQUBO2QAIA, built on relevant toolkits, that support dimensionality reduction of higher-order Ising models for objective functions with complex coefficients;

(7) Numerical computation and visualization: auxiliary programs for computing and comparing various performance metrics, as well as generating plots.

## 4 Experiments and Analysis

### 4.1 Scoring Criteria and Test Environment

#### 4.1.1 Scoring Criteria

According to the requirements of the 7th National Quantum Computing Hackathon, the submitted algorithm must output an optimized phase sequence $\{\alpha_n\}$ and amplitude sequence $\{\beta_n\}$ for each test case, and its performance is evaluated by a scoring function. The core of the scoring rule is to balance three metrics: sidelobe-suppression capability, main-beam-width control accuracy, and main-beam pointing accuracy. The specific scoring mechanism is as follows:

$$y_i = 1000 - 100a - 80b - 20c$$

Here $y_i$ is the score of a single test case, with 1000 as the full score; $a$ is the penalty term for wide-angle sidelobe suppression; $b$ is the penalty term for main-beam width; and $c$ is the penalty term for near-main-lobe sidelobe suppression. The detailed rules for penalty calculation are:

$$a = max\left\{15 + max\left\{10\lg\frac{|F(\theta)|^2}{max|F(\theta)|^2}\right\}, 0\right\}, \theta \in [0, \theta_0 - 30°) \cup (\theta_0 + 30°, 180°]$$

$$b = max\{W - 6°, 0°\}$$

$$c = max\left\{10\lg\frac{|F(\theta)|^2}{max|F(\theta)|^2} + 30\right\}, \theta \in [\theta_0 - 30°, \theta_1] \cup [\theta_2, \theta_0 + 30°]$$

In addition, if the main-beam pointing deviation exceeds the limit, namely $|\theta - \theta_0| \leq 1°$, or if the optimization time for a single case exceeds 90 seconds, the score is directly set to $y_i = 0$. The final result is the arithmetic mean of the scores over all test cases:

$$y = \frac{1}{N_{case}} \sum_{i=1}^{N_{case}} y_i$$

### 4.1.2 Test Environment

The software development environment is built on the Python programming language and does not use any commercial optimization or numerical-computation modules prohibited by the competition rules, strictly following the parameter ranges and data-type requirements defined in the competition specification. The core implementation in this work was developed using the MindSpore Quantum framework [15], together with standard open-source scientific-computing tools in Python. These software tools provide a unified environment for algorithm implementation, numerical simulation, and performance evaluation while remaining fully compliant with the competition requirements. The local hardware environment is an ordinary computer and does not require any special acceleration devices. All algorithms and techniques proposed in this paper can be implemented and executed on standard general-purpose computing equipment. The online test environment uses the Huawei Cloud resources specified by the competition, with instance type memory-optimized m7, specification m7.4xlarge.8, and 2U4G.

Following the competition's description of the dataset, a self-developed data-generation script was used to produce local test cases that satisfy the specification. Specifically, a random-seed mechanism was used to control the generation process, ensuring that each test case contains three core parameters: the target main-beam pointing angle $\theta_0$, which is a floating-point number in the range 45 degrees to 134 degrees; the number of phase-encoding bits, an integer randomly selected from {1,2,3,4}; and the amplitude-optimization flag, a random Boolean value of True or False. The data structure, numerical ranges, and distribution characteristics are fully consistent with the technical requirements of the competition problem. The code required to reproduce the results reported in this paper are publicly available at https://gitee.com/mindspore/mindquantum/tree/research/hackathon/hackathon2025/qaia/new.

### 4.2 Algorithm Performance Tests

#### 4.2.1 Comparison Between Baseline Algorithms and the Hybrid-Framework Algorithm

As can be seen from Fig. 3, in experimental tests on the beamforming optimization problem, the algorithm proposed in this paper, namely the hybrid-framework algorithm, shows performance advantages over both baseline algorithms, i.e., the classical gradient algorithm and the quantum-inspired algorithm.

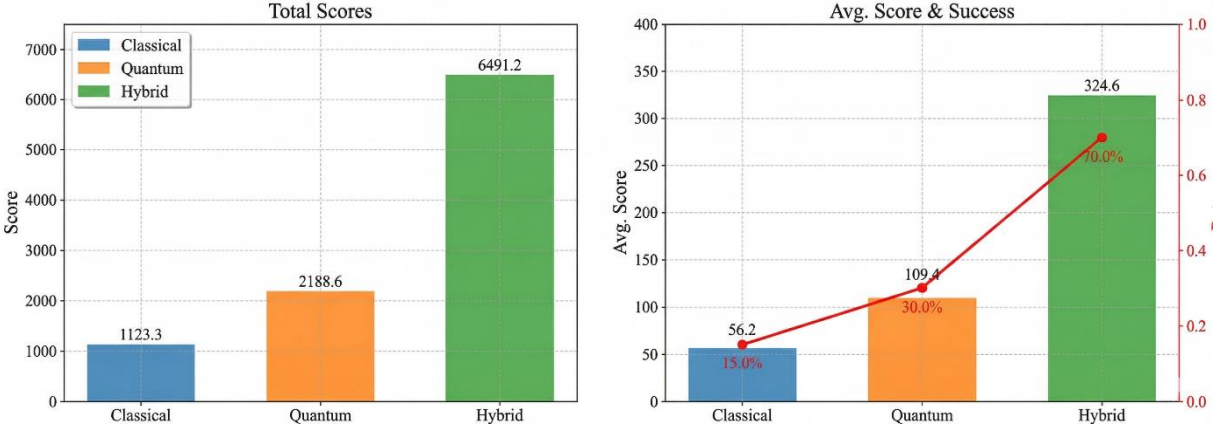

**Fig. 3 Algorithm Scores and Optimization Success Rates**

The left panel compares total scores. The hybrid-framework algorithm achieves a cumulative score of 6491.2, which is 197 percent higher than the baseline quantum-inspired algorithm with 2188.6 points and 479 percent higher than the baseline classical gradient algorithm with 1123.3 points.

The right panel compares average score and success rate, further reflecting the performance advantage of the proposed hybrid-framework algorithm over the two baseline algorithms. The average score of the hybrid framework is 324.6, compared with 109.4 for the baseline quantum-inspired algorithm and 61.8 for the baseline classical gradient algorithm. This indicates that, in scenarios where valid solutions are found, the hybrid framework delivers much higher solution quality than the baseline methods. More importantly, the success-rate metric shows that the proposed hybrid-framework optimization algorithm obtains valid solutions in 70 percent of the test cases, far exceeding the 30 percent success rate of the baseline quantum-inspired algorithm and the 15 percent success rate of the baseline classical gradient algorithm. This gap shows that by combining two optimization strategies, the hybrid framework greatly overcomes the instability of the quantum-inspired algorithm, as will also be seen in later experiments, while also overcoming the classical gradient algorithm's tendency to get trapped in local optima. This improvement in robustness implies that deploying the hybrid-framework algorithm in real MIMO systems could greatly reduce the frequency of system restarts for recomputation.

Fig. 4 shows a scatter plot comparing the scores of the proposed hybrid-framework algorithm with those of the two baseline algorithms.

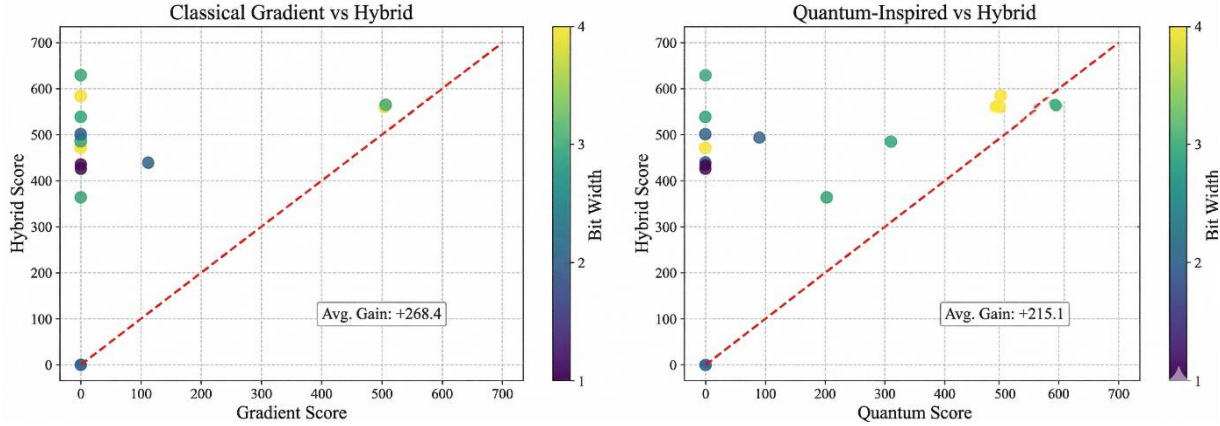

Fig. 4 Scatter Plot of Algorithm Scores

In the left scatter plot, the performance of the baseline classical gradient algorithm is compared with that of the proposed hybrid-framework algorithm. The distribution shows that most points lie in the upper-left region relative to the diagonal, which directly reflects the performance advantage of the hybrid framework under the same test scenarios. First, the hybrid-framework algorithm outperforms the classical gradient algorithm in every test case, with an average improvement of 268.4 points. Second, among the 17 cases in which the classical gradient algorithm scored zero, the hybrid-framework algorithm successfully found valid solutions in 11 cases, clearly demonstrating its advantage in scenarios where the traditional gradient method fails completely. Moreover, in the two cases where the classical gradient algorithm achieved high scores above 500, the hybrid framework still maintained stable performance, with a minimum score of 560.4 and a maximum score of 564.8. This indicates that even when the baseline method performs well, the hybrid framework can maintain stable output without performance degradation.

Similarly, the right scatter plot shows that the hybrid-framework algorithm scores lower than the quantum-inspired algorithm in only one test case, while achieving an average improvement of 215.1 points. Particularly noteworthy are the three points farthest from the diagonal, for which the quantum-inspired algorithm scored zero while the hybrid-framework algorithm scored above 500 in all cases, with a maximum of 629.3. This reveals the unique advantage of the hybrid framework in handling cases where the quantum-inspired algorithm fails completely. In addition, the coloring by bit width shows that as the bit number increases, from light yellow to dark purple, the advantage of the hybrid-framework algorithm becomes increasingly pronounced. Especially in 3-bit and 4-bit scenarios, both the classical gradient algorithm and the quantum-inspired algorithm score lower than the hybrid framework, reflecting the high performance of the proposed hybrid approach in high-dimensional scenarios.

**4.2.2 Comparison Between the Baseline Phase-Encoding Scheme and the Proposed Phase-Encoding Scheme**

Figure 5 presents the performance comparison between the baseline phase-encoding scheme and the proposed phase-encoding scheme.

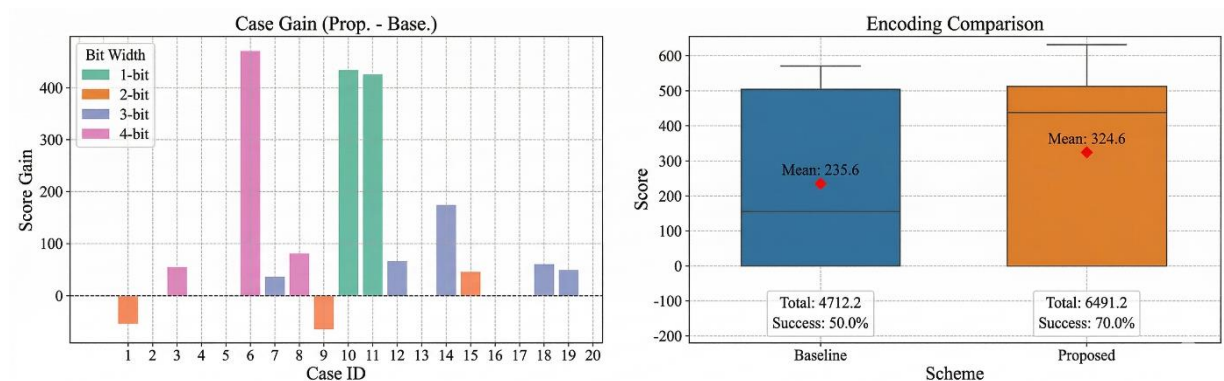

**Fig. 5 Performance Comparison of Different Encoding Schemes**

The left panel shows the score difference for each test case, with the horizontal axis representing the case ID and the vertical axis representing the score of the proposed phase-encoding scheme minus that of the baseline phase-encoding scheme. Values above zero indicate that the proposed scheme is better, while values below zero indicate that the baseline scheme is better. Several observations can be made. First, in the vast majority of cases, the score of the proposed encoding scheme is no lower than that of the baseline scheme. Second, when cases with different bit widths are examined by color, there are two 2-bit cases in which the baseline encoding scheme performs better. However, it should be noted that in 2-bit scenarios the two schemes are actually equivalent, so the apparent advantage of the baseline in those two cases is merely due to random chance. As the number of cases increases, this randomness will inevitably diminish. In addition, the proposed encoding scheme has clear advantages in 3-bit and 4-bit cases, which reflects its stronger advantage in high-bit-width scenarios. The reason is that the baseline phase-encoding scheme uses a fixed 2-bit encoding, whereas the proposed scheme uses the variable-bit Gray-odd-combination encoding described earlier.

The right panel is a boxplot comparing the overall performance of the two encoding schemes. It can be seen that the total score of the proposed encoding scheme reaches 6491.2, exceeding the baseline score of 4712.2 by nearly 38 percent. There is also a significant difference in the success-rate metric: the proposed encoding scheme successfully outputs valid solutions in 14 cases, corresponding to a success rate of 70 percent, whereas the baseline encoding scheme succeeds in 10 cases, corresponding to a success rate of 50 percent, a gap of 30 percentage points. This success-rate advantage is also reflected in the boxplot distribution. The median of the proposed encoding scheme is around 430, while the median of the baseline scheme is around 150, indicating a clear overall performance advantage of the proposed encoding scheme.

Figure 6 shows the performance of the baseline encoding scheme and the proposed encoding scheme under different scenarios.

The left panel compares performance under different encoding bit settings. As shown, in the low-dimensional 1-bit scenario, the proposed scheme achieves an average score of 122.9, whereas the baseline scheme fails completely and scores zero. The proposed encoding scheme also has obvious advantages in high-bit scenarios: both in 3-bit and 4-bit cases, its average score is significantly higher than that of the baseline encoding. Specifically, it is higher by 18 percentage points in 3-bit scenarios and 60 percentage points in 4-bit scenarios. The only case in which the proposed scheme lags is the 2-bit scenario, but as noted earlier, the two schemes are actually equivalent there, and the observed gap comes from random chance. It is also worth noting that in 3-bit and especially 4-bit cases, the score variance of the proposed encoding scheme is noticeably smaller than that of the baseline scheme, indicating better stability in high-bit scenarios.

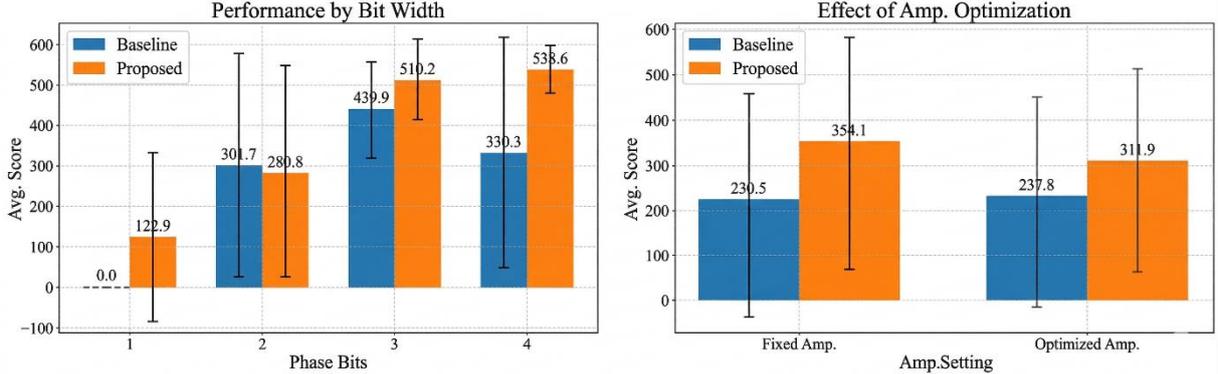

Fig. 6 Performance Comparison Under Different Scenarios

The right panel compares performance with and without amplitude optimization. It can be seen that the proposed encoding scheme outperforms the baseline scheme regardless of whether the amplitude is fixed or optimized. In the fixed-amplitude scenario, the average score of the proposed encoding scheme is nearly 100 points higher. In the amplitude-optimizable scenario, the baseline scheme partially recovers some of its disadvantage, but its average score still lags behind the proposed scheme by about 20 percent. These comparisons indicate that the proposed encoding scheme has better adaptability and stability across different scenarios.

### 4.2.3 Comparison Among No Amplitude Encoding, Equal-Value Amplitude Encoding, and the Proposed Amplitude-Encoding Scheme

Figure 7 shows the main statistical indicators of the three schemes, namely no amplitude encoding, equal-value amplitude encoding, and the proposed amplitude encoding, after excluding cases with zero scores. The figure demonstrates the performance advantage of the proposed scheme.

First, in terms of the mean, the proposed scheme achieves an average score of 499.3, which is 32.9 percent higher than the 375.8 of the no-amplitude-encoding scheme and 29.8 percent higher than the 384.8 of the equal-value amplitude-encoding scheme, reflecting the effectiveness of the proposed amplitude encoding. Second, in terms of the median, the proposed scheme reaches 493.8, which is 19.3 percent higher than the 413.8 of the equal-value amplitude-encoding scheme and more clearly demonstrates a stable advantage in more than half of the test scenarios, proving the algorithm's strong robustness to environmental-parameter changes. Third, extreme-value analysis further shows that the maximum value of 629.3 is 7.0 percent higher than the best solution of 588.1 without amplitude encoding, while the minimum value of 363.4 is roughly four times the 71.9 of the no-amplitude-encoding scheme, verifying the adaptability of the algorithm under different scenarios. In addition, the standard deviation of 71.5 is only 56 percent of the 126.9 for equal-value amplitude encoding and 48 percent of the 147.7 for no amplitude encoding, and this lower fluctuation further confirms the stability of the proposed amplitude-encoding scheme. For example, when the quantization bits of the main-beam direction are limited, using the proposed scheme to encode amplitude at higher resolution allows finer amplitude adjustment and better suppression of sidelobe energy leakage.

Figure 8 shows the distribution of performance improvement brought by the proposed amplitude-encoding scheme and helps explain the mechanism behind the technical advantage. First, compared with the no-amplitude-encoding scheme, the proposed scheme yields an average improvement rate of 90.2 percent. The improvement curve exhibits a single-peak long-tail characteristic: the main peak is around 20 percent, indicating stable small-to-medium gains under common environments, while the long tail concentrated in the 300 to 800 percent region reflects that in some scenarios the proposed scheme can provide very substantial improvements in algorithm performance. Second, compared with equal-value amplitude encoding, the improvement distribution takes a bimodal shape: the main peak is still around 20 percent, the average improvement is 53.1 percent, and the concentration is high. This indicates stable small-to-medium performance gains under common environments. The secondary peak is around 450 percent, again showing that there exist amplitude values that equal-value encoding cannot effectively optimize but the proposed scheme can reach. Finally, note that both the no-amplitude-encoding and equal-value-amplitude-encoding comparisons still have a large area in the negative-improvement region, but this is mainly due to the smoothing effect of the data under the condition that the main peak is around 20 percent. Figure 7 shows that the actual scores of the proposed scheme are in fact never lower than those of the comparison schemes.

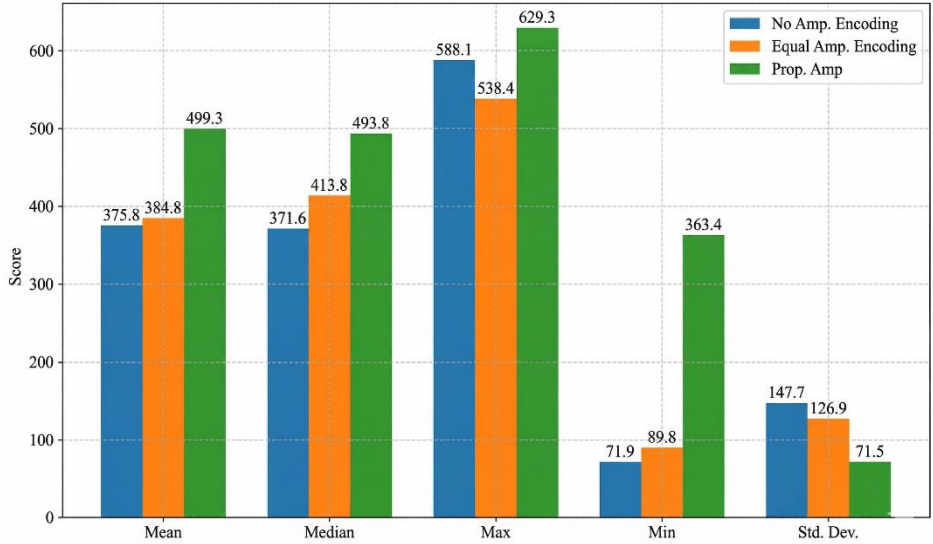

**Fig. 7 Comparison of Main Statistical Indicators of Different Amplitude-Encoding Schemes**

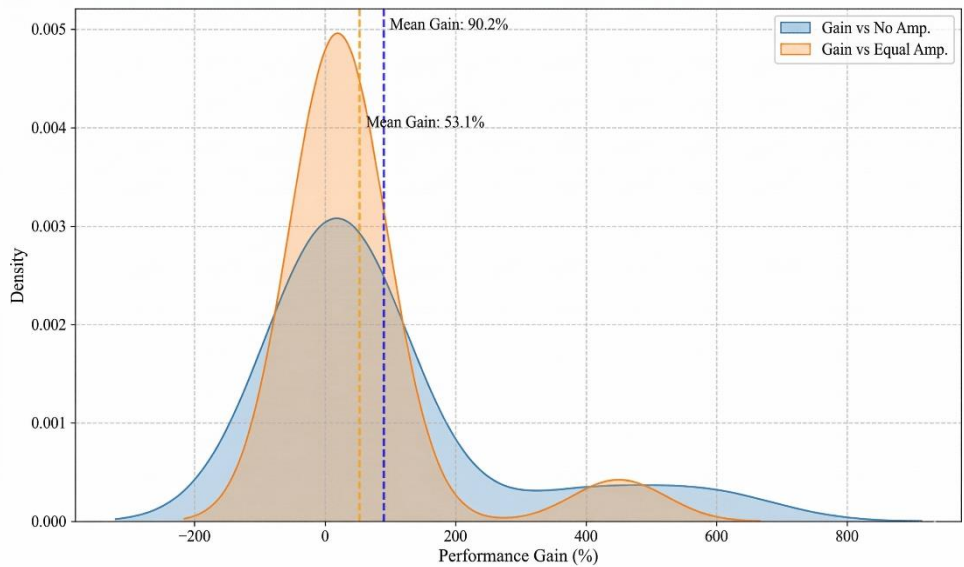

**Fig. 8 Distribution of Performance Improvement Brought by the Proposed Amplitude-Encoding Scheme**

### 4.2.4 Comparison Between a Single Quantum-Inspired Algorithm and the Rainbow Quantum-Inspired Algorithm

Figure 9 compares a single quantum-inspired algorithm, using the BSB solver in the baseline algorithm as the reference and referred to below as the single-heuristic scheme, with the proposed rainbow quantum-inspired algorithm combined with candidate-solution refinement, referred to below as the rainbow-refinement scheme.

The left panel shows the score difference for each test case, with the horizontal axis representing the case ID and the vertical axis representing the score of the rainbow-refinement scheme minus that of the single-heuristic scheme. Values above zero indicate that the rainbow-refinement scheme is better, while values below zero indicate that the single-heuristic scheme is better. The figure shows first that the score of the rainbow-refinement scheme is no lower than that of the single-heuristic scheme in all cases. Looking at cases with different bit widths, indicated by different colors, the rainbow-refinement scheme is no worse than the single-heuristic scheme under any bit setting, and no point crosses below zero. This demonstrates that compared with the single-heuristic scheme, the rainbow-refinement scheme produces solutions of consistently higher quality and stability. It is also worth noting that the advantage of the rainbow-refinement scheme becomes particularly clear in 2-bit, 3-bit, and 4-bit cases, reflecting its stronger superiority in high-bit encoding scenarios.

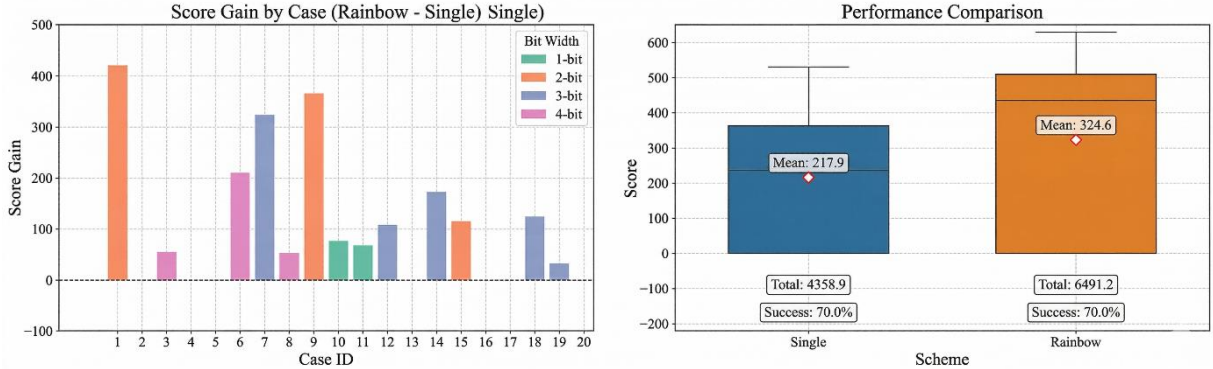

Fig. 9 Performance Comparison of Different Quantum-Inspired Algorithm Schemes

The right panel is a boxplot comparing the overall performance of the two quantum-inspired schemes. The total score of the rainbow-refinement scheme reaches 6491.2, exceeding the single-heuristic scheme by more than 49 percent over the 4358.9 score of the latter. Although the success-rate metric is the same for both schemes, namely 70 percent, the performance advantage of the rainbow-refinement scheme is evident in the boxplot distribution: its median is around 420, whereas the median of the single-heuristic scheme is below 250. At the same time, the average score of the rainbow-refinement scheme is 324.6, which is also substantially higher than the 217.9 of the single-heuristic scheme. All these indicators clearly demonstrate the performance advantage of the rainbow-refinement scheme over the single-heuristic scheme.

Figure 10 shows the performance of the single-heuristic scheme and the rainbow-refinement scheme under different scenarios.

The left panel compares performance under different encoding bit settings. It can be seen that in the low-dimensional 1-bit scenario, the rainbow-refinement scheme achieves an average score of 122.9, whereas the single-heuristic scheme achieves 102.1, so the rainbow-refinement scheme is only slightly better. In contrast, the advantage of the rainbow-refinement scheme is much more pronounced in high-bit scenarios. As shown in the figure, in the 2-bit cases the rainbow-refinement scheme has the largest advantage, exceeding the single-heuristic scheme by more than 170 percent. In 3-bit and 4-bit scenarios, the advantage becomes smaller but still remains around 20 to 30 percent. It is also worth noting that in these higher-bit scenarios, especially 3-bit and 4-bit, the score variance of the rainbow-refinement scheme is clearly smaller, indicating better performance stability than the single-heuristic scheme.

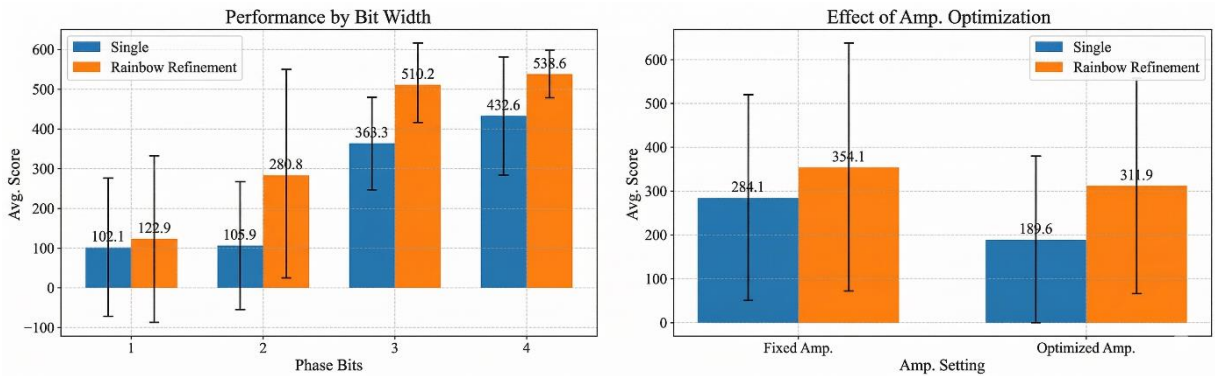

Fig. 10 Performance Comparison Under Different Scenarios

The right panel compares performance with and without amplitude optimization. It can be seen that the rainbow-refinement scheme outperforms the single-heuristic scheme regardless of whether the amplitude is fixed or optimizable. In the optimizable-amplitude scenario, the average score of the rainbow-refinement scheme is 122.3 points higher; in the fixed-amplitude scenario, the gap between the two schemes narrows, but the average score of the single-heuristic scheme still lags behind the rainbow-refinement scheme by 25 percentage points. These comparisons show that the rainbow-refinement scheme has better environmental adaptability.

In fact, the performance gap between a single quantum-inspired algorithm and the rainbow quantum-inspired algorithm combined with candidate-solution refinement is not unique to the BSB case. Similar effects can be observed when BSB is replaced by other quantum-inspired algorithms. Table 1 gives a score comparison between various single quantum-inspired algorithms and the proposed rainbow-refinement scheme.

**Table 1 Comparison of Scores Between a Single Quantum-Inspired Algorithm and the Proposed Rainbow-Refinement Scheme**

| Algorithm | Score | Performance Difference |
| --- | --- | --- |
| Rainbow-Refinement Scheme | 6491.2 | - |
| BSB | 4358.9 | -32.8% |
| DSB | 4383.3 | -32.5% |
| SimCIM | 1341.1 | -79.3% |
| LQA | 4226.3 | -34.9% |
| CAC | 3982.2 | -38.7% |
| CFC | 4065.4 | -37.4% |
| NMFA | 4277.5 | -34.1% |

### 4.2.5 Contribution of Quantum-Inspired Algorithms and Classical Gradient Algorithms in the Hybrid Framework

Within the hybrid framework, quantum-inspired algorithms and classical gradient algorithms each play complementary roles.

Based on the contribution ratio of each module in the workflow to the overall score, Fig. 11 plots the contributions of the quantum-inspired and classical gradient algorithms in the hybrid framework. The figure shows that the quantum-inspired algorithm plays the dominant role throughout the optimization process. In particular, during phase optimization its contribution reaches about 78 percent. The subsequent amplitude optimization based on the quantum-inspired algorithm also contributes about 14 percent. This indicates that the hybrid framework successfully develops and exploits quantum-inspired algorithms. The figure also shows that although the contribution of the classical method is relatively small, its overall contribution is still close to 10 percent, serving as a useful supplement to the quantum-inspired algorithm. In practice, the contribution ratio is not fixed. Depending on the test case and application scenario, different parts of the hybrid framework may have the opportunity to play the dominant optimization role.

Overall, by organically combining quantum discrete exploration of phase and amplitude with classical continuous fine tuning, the hybrid framework retains the global escape capability of quantum-inspired algorithms while also inheriting the precision-refinement advantage of classical gradient methods, thereby improving both performance and robustness. This design idea, together with the contribution distribution revealed by the experimental data, provides a useful reference for reasonably allocating quantum and classical resources in more hybrid optimization scenarios in the future.

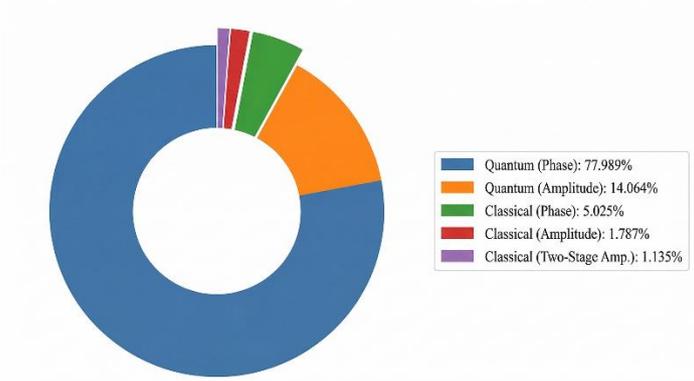

Fig. 11 Contribution of Quantum and Classical Algorithms in the Hybrid Framework

## 4.3 Overall Technical Evolution and Online Scores

Figure 12 presents the technical evolution map of this study, outlining the research trajectory of the algorithm and the path of performance improvement from the baseline algorithm to the current solution.

The first attempt at the official competition example program used the single quantum-inspired algorithm provided by the organizers. However, because its default batch size is 1 and the solution quality is unstable, the baseline algorithm was chosen to be another program provided by the organizers, namely the classical gradient-optimization algorithm. The online test score of this baseline algorithm was 229.8375.

The first stage of performance improvement came from constructing the proposed hybrid optimization architecture that combines quantum-inspired and classical algorithms. By establishing a dual-track coordination mechanism between the BSB quantum-inspired method and gradient optimization, the solution score increased sharply to 356.0597, a 54.9 percent improvement. This was achieved by preserving the advantage of discrete-phase optimization in the quantum-inspired path, releasing the amplitude degree of freedom in the classical gradient path, improving the batch-size hyperparameter, and introducing multiple candidate solutions.

The key innovation of this stage lies in building the hybrid architecture and the matching implementation details, such as batch-size adjustment and candidate-solution fusion and evaluation. These designs effectively overcome the disconnected optimization of discrete and continuous variables in traditional methods and bridge the originally separated islands of the solution space. However, at this stage the quantum-inspired path and the classical gradient path remained relatively independent and were not deeply fused, especially because the quantum-inspired path had not yet incorporated gradient-based amplitude optimization. This stage is labeled version 1.0 of the hybrid framework in the figure.

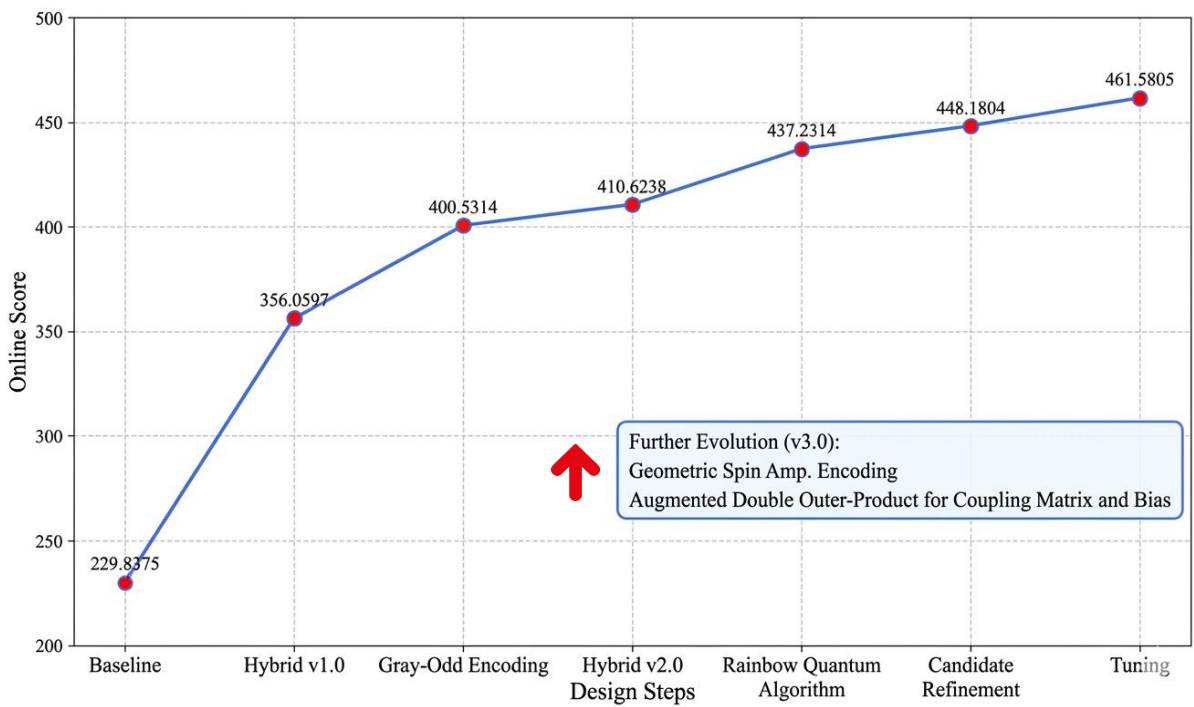

**Fig. 12 Technical Evolution and Scores**

The second stage of performance improvement came from the design of the phase-encoding scheme. The baseline algorithm uses a fixed 2-bit optimization, which performs acceptably in 1-bit and 2-bit test cases. However, in 3-bit and 4-bit scenarios, the fixed encoding suffers from insufficient precision. In addition, although the competition reference paper provided an encoding scheme, directly using it in 3-bit and 4-bit cases either leads to a higher-order Ising model or causes a surge in spin variables after dimensionality reduction. Therefore, we designed an encoding scheme based on Gray-odd-combination encoding together with several auxiliary techniques. Through Gray-code transformation and odd-subset feature extraction, the Gray-odd-combination encoding constructs a phase-mapping system with adaptive dimensionality. Combined with the double outer-product method, it makes it very convenient to construct phases with arbitrary bit widths. In terms of online scores, opening up the freedom of multi-bit encoding enabled the score to exceed 400 for the first time. After several rounds of further fine tuning to optimize multi-bit scenarios and the introduction of gradient-based amplitude optimization into the quantum-inspired path, the cooperation between the quantum and classical algorithms became tighter and more coordinated. The hybrid framework thus advanced to stage 2.0, and the online score also reached 410.6238.

The third stage of performance improvement came from the implementation of the rainbow quantum-inspired algorithm and candidate-solution refinement. This method integrates seven optimizers based on different mathematical and physical principles. BSB simulates nonlinear resonator dynamics for coarse-grained exploration. DSB optimizes the branch structure of the solution space during discrete bifurcation. SimCIM performs pulse iteration based on an optical interference model to expand solution coverage. LQA builds annealing optimization over multiple local regions to improve solution diversity and convergence stability. NMFA uses mean-field annealing with noise perturbations to enhance the ability to jump and escape local optima. CAC introduces controlled chaos into the energy landscape for global probing of the solution space. CFC uses a feedback loop to generate perturbations that help the algorithm escape local extrema. All algorithms run in parallel under a unified Hamiltonian framework, forming full-spectrum coverage of the solution space. In addition, the supporting candidate-solution refinement mechanism uses hierarchical clustering based on distance and similarity to greatly compress the candidate pool while preserving high quality and diversity, enabling the hybrid framework to output high-quality candidate solutions more efficiently and more stably. After this stage of technical evolution, the online score increased to 448.1804.

The final stage of performance improvement came from the application of the double outer-product method and parameter tuning. Traditional construction of the coupling matrix depends on time-consuming element-wise operations and takes away time from other optimization tasks. The double outer-product method simplifies the algorithmic flow through tensor operations: the first outer product generates the matrix of field-coefficient products, and the second outer product directly outputs the target coupling matrix after taking the real part, greatly reducing computation time. This advantage is particularly prominent in high-bit scenarios. In addition, parameters and hyperparameters were optimized through a combination of Bayesian optimization, grid search, and manual tuning to determine appropriate numerical values. Through the combined effect of these techniques and efforts, the online score of the algorithm reached 461.58, more than doubling the score of the baseline algorithm.

It should be added that after the online scoring website was closed, that is, after the preliminary round ended, the research work continued to advance. We further designed a geometric amplitude-encoding scheme and introduced the augmented outer-product method to construct the coupling matrix and bias vector of the amplitude objective function. This allows the rainbow quantum-inspired algorithm to be applied in both phase optimization and amplitude-value optimization. Accordingly, amplitude optimization was divided into two stages: coarse tuning by the quantum-inspired algorithm, which is a discrete optimization process used to escape local optima, and fine refinement by the classical gradient algorithm, which is a continuous optimization process used to obtain precise extrema. This further improves algorithm performance and the overall degree of integration of the hybrid framework. We refer to the current stage of technical evolution as version 3.0 of the hybrid framework.

## 5 Conclusion and Outlook

This paper proposes an algorithm based on a quantum-classical hybrid optimization framework for large-scale antenna-array beamforming, aiming to efficiently solve the difficult joint optimization of discrete phase and continuous amplitude. By combining quantum-inspired algorithms with classical gradient-optimization methods, the algorithm demonstrates significant advantages in several aspects:

(1) Algorithmic innovation and performance improvement

The proposed hybrid optimization framework effectively integrates the efficient search capability of quantum-inspired algorithms in the discrete-phase space with the fine adjustment capability of classical gradient optimization in the continuous-amplitude space, overcoming the limitations of traditional single-paradigm methods in mixed-variable optimization. Experimental results show that, under the constraints on sidelobe intensity near the main-beam direction, sidelobe intensity over a wider range, main-beam width, and optimization time, the online performance score of the algorithm reaches 461.58, nearly doubling that of the baseline algorithm.

(2) Design of the phase-encoding scheme

A phase-encoding scheme based on Gray codes and odd-combination products is designed, which improves encoding robustness, avoids the exponential growth of the complexity of higher-order Ising models, and provides a universal theoretical basis for phase optimization at arbitrary bit precision. This encoding scheme shows clear advantages in 3-bit and 4-bit cases and significantly improves the performance and stability of the algorithm.

(3) Design of the amplitude-encoding scheme

An amplitude-encoding scheme based on geometric combinations of spin variables is designed. Without increasing the model order, it realizes high-resolution representation of discrete amplitude values and provides support for the efficient application of quantum-inspired algorithms to amplitude optimization.

(4) Rainbow quantum-inspired algorithm and adaptive deduplication optimization scheme

The proposed rainbow quantum-inspired algorithm effectively improves solution diversity and quality by integrating multiple optimizers to explore the solution space in parallel and combining them with an adaptive deduplication mechanism. This scheme shows significant advantages in high-bit encoding scenarios and further improves the performance and stability of the algorithm.

(5) Constructing the objective-function coupling matrix and bias vector by the double outer-product method and the augmented double outer-product method

Tensor outer products and the augmented double outer-product method are used to construct the coupling matrix and bias vector of the objective function, thereby improving numerical precision and computational efficiency. This construction method is also naturally compatible with encoding operations at arbitrary bit widths, reducing program-design difficulty and improving program scalability.

(6) Exploration of higher-order Ising-model solving

The difficulty of solving higher-order Ising models is analyzed, and the possibility, advantages, and disadvantages of exchanging more spin-variable bits for lower dimensionality are explored, providing reference ideas for solving beamforming optimization problems with higher-order encodings.

Although the algorithm proposed in this paper achieves significant performance improvement for large-scale antenna-array beamforming, there is still room for further enhancement. For example, exploring the possibility of synchronous optimization of phase and amplitude would be a very interesting direction for future research. How to jointly encode phase and amplitude, how to avoid higher-order Ising models or perform effective dimensionality reduction through some technique, and related issues all remain to be developed in this direction. In addition, future extensible directions include adaptive encoding of the main-beam angle, algorithm parallelization, hardware implementation, multi-scenario and multi-objective adaptability, and integration with other communication technologies.

In summary, the research in this paper provides an efficient and robust reference solution for antenna-array beamforming in communication systems. It also opens a new technical route for integrating quantum-inspired and classical algorithms and applying them in the communication field. Future work will continue along these directions to promote the further development and application of quantum-inspired algorithms and beamforming technology.

## Acknowledgement

This work was sponsored by CPS-Yangtze Delta Region Industrial Innovation Center of Quantum and Information Technology-MindSpore Quantum Open Fund.